\documentclass[aps,prb,preprint,epsfig,superscriptaddress,showpacs,amsmath,amsfonts,amssymb,floatfix]{revtex4-1}
\usepackage{graphicx}
\usepackage{dcolumn}
\usepackage{bm}
\usepackage{epstopdf}
\usepackage{dsfont}

\newcommand{\bra}[1]{\left\langle #1\right|}
\newcommand{\ket}[1]{\left| #1\right\rangle}

\usepackage[usenames,dvipsnames]{color}

\begin{document}

\title{Topological Effects in Chiral Symmetric Driven Systems}

\author{Derek Y.H. Ho \footnote{Current address: Graphene Research Center and Department of Physics, National University of Singapore, 117546, Singapore}}
\affiliation{Department of Physics and Center for Computational
Science and Engineering, National University of Singapore, 117542,
Singapore}
\author{Jiangbin Gong}
\email{phygj@nus.edu.sg}
\affiliation{Department of Physics and Center for Computational
Science and Engineering, National University of Singapore, 117542,
Singapore}
\affiliation{NUS Graduate School for Integrative Sciences
and Engineering, Singapore 117597, Singapore}

\begin{abstract}

Recent years have seen a strong interest in topological effects within periodically driven systems. In this work, we explore topological effects in two closely related 2-dimensional driven systems described by Floquet operators possessing chiral symmetry (CS). Our numerical and analytical results suggest the following. Firstly, the CS is associated with the existence of the anomalous counter-propagating (ACP) modes reported recently.
Specifically, we show that a particular form of CS protects the ACP modes occurring at quasienergies of $\pm \pi$. We also find that these modes are only present along selected boundaries, suggesting that they are a weak topological effect. Secondly, we find that CS can give rise to protected $0$ and $\pi$ quasienergy modes, and that the number of these modes may increase without bound as we tune up certain system parameters. Like the ACP modes, these $0$ and $\pi$ modes also appear only along selected boundaries and thus appear to be a weak topological effect. To our knowledge, this work represents the first detailed study of weak topological effects in periodically driven systems. Our findings add to the still-growing knowledge on driven topological systems.

\end{abstract}

\date{\today}
\maketitle

\section{Introduction}

Topological effects in periodically driven systems are by now a subject of considerable theoretical and experimental interest \cite{ZhaoKickedHall,AsbothBBC,AsbothBBC1DSys,KitagawaQwalk,KitagawaTopoChar,JiangPRL,HoGongPRL2012,LongWenEPJB,FTINatPhys,AsbothSymmetries,
KitagawaNatComm,BLGraphenePRB,dora,NatureFTI,RudnerPRX,ThakurathiPRB,PodolskyMFTI,ManyMajoranasPRB,
FloquetGraphenePRB,PhotonicQWPRL,PlateroPRB2013,PlateroPRL2013,DisorderFTI,GoldmanPRX,FloquetSpinHall,ZhaoLatest}.
The main reason for this interest is that driving fields offer an easy way of tuning a system's topological properties, unlike the case for static systems in which their topological properties are for most intents and purposes fixed during the fabrication process. This fact was first theoretically demonstrated in Ref. \cite{FTINatPhys}, in which a system was tuned from being topologically trivial to non-trivial by means of a driving field, forming a Floquet topological insulator (FTI). Recently, this effect has also been theoretically demonstrated in graphene \cite{FloquetGraphenePRB}. There are also various interesting effects which are peculiar to driven systems, of which we name only a few for brevity (see Ref. \cite{dora} for a review). Firstly, driven systems can host two types of edge modes with zero group velocity. These are the $0$ or $\pi$ quasienergy edge modes \cite{KitagawaQwalk,KitagawaNatComm,AsbothBBC,AsbothBBC1DSys}, whereas static systems can only give rise to zero-energy edge modes. Secondly, driven systems allow for the generation of Floquet Majorana modes \cite{JiangPRL} which are described by different invariants \cite{ThakurathiPRB} than their static counterparts and may in theory be generated in large number simply by increasing the period of the driving field \cite{ManyMajoranasPRB}. Thirdly, driving fields have been proposed as a means of achieving a semimetal-insulator phase transition in graphene \cite{PlateroPRB2013}. To understand these driving-induced effects, general theoretical frameworks for solving driven lattice systems have been proposed in Refs. \cite{PlateroPRL2013,GoldmanPRX}. Fourthly, driving fields have also been found to induce anomalous counter-propagating (ACP) chiral edge modes in lattice systems \cite{ZhaoKickedHall,ZhaoLatest}, which are currently not well understood theoretically (see below for details). Most recently, a driving field-induced spin Hall effect has been theoretically proposed \cite{FloquetSpinHall}.

Along a separate vein, the subject of weak topological insulators (WTIs) is also an area of ongoing research activity \cite{Kobayashi-2013PRL,Yoshimura-2013PRB,Hatsugai-2013JpnSoc,Shen-2014arXiv,Hatsugai-2014arXiv}. WTIs are mostly spoken of in the context of three-dimensional (3D) systems \cite{Fu-Kane-Mele-2007PRL,Moore-Balents-2007PRB}. These phases are ‘weak’ for two reasons. Firstly, the edge modes of WTIs only exist along certain boundaries (i.e., it depends on the shape one cuts out from an infinite lattice to obtain a finite sample and also on which boundary of the sample one looks at), in contrast to the situation for strong topological insulators (STI), in which edge modes always exist on the boundary regardless of its shape and direction. Secondly, WTIs are in general not stable in the same way that STIs are, because they are associated with invariants in a lower dimensionality than that of the physical system.
It was first thought that because coupling the two-dimensional (2D) layers of a 3D WTI system in pairs renders them topologically trivial\cite{Fu-Kane-Mele-2007PRL,Ringel-2012PRB}, WTIs are no more interesting than topologically trivial insulators. However, it was pointed out in several papers \cite{Ringel-2012PRB,Mong-2012PRL, Kobayashi-2013PRL} that random disorder, which is unlike the coherent pair-wise coupling mentioned above, is insufficient for localizing all the edge states. Also, several papers showed that the lower-dimensional topological effects can manifest as protected conducting channels at dislocations \cite{Ran-2009NatPhys} or `terrace-structures' \cite{Imura-2011PRB, Yoshimura-2013PRB} in 3D lattices. In other words, WTIs display topological effects because the lower-dimensional topological non-triviality survives even in the higher dimension. Recently, there has been an interest in WTIs in the context of 2D systems \cite{Shen-2014arXiv,Hatsugai-2013JpnSoc, Hatsugai-2014arXiv}. Refs. \cite{Hatsugai-2013JpnSoc, Hatsugai-2014arXiv} introduced topological numbers which demonstrate the bulk-boundary correspondence of 2D WTIs, while Ref. \cite{Shen-2014arXiv} studied the 1-dimensional (1D)-2D transition in topological behavior of a square lattice as it is gradually built up from stacking 1D chains atop one another. Along another line of work, it is also known that chiral symmetry gives rise to interesting topological zero modes and Dirac points in graphene.\cite{Hatsu-topo-graphene,Hatsu-CS-graphene,Koshino2014}

The present work studies weak topological effects in chiral symmetric driven systems, making it relevant to all the three themes above- namely FTIs, WTIs and chiral symmetry. We study these effects in the context of two topologically non-trivial driven 2D models. These are lattice versions of
the kicked Harper model (KHM) \cite{zas,FastDelocalizatn,MetamorphosisPRL} (this was regarded as a kicked quantum Hall system in Ref.~\cite{ZhaoKickedHall}) and an on-resonance double-kicked rotor (ORDKR) model \cite{Monteiro,Gong2008proposalPRA}.
We chose to study these two models because their eigenstates are related by a precise 1-1 correspondence which ensures that their Chern numbers (when computable) must be equal \cite{KHMvsORDKR} (see Sec. III for details). This guaranteed equality of Chern numbers suggests that any differences in topological edge state behaviour seen in the two models is not directly related with the Chern numbers, making this a good opportunity to study weak topological behaviour. In this work, we study these weak topological properties analytically and numerically. It is our hope that this work motivates further investigations into weak topological effects in driven systems, an area which is still unexplored and might thus contain new physics.

We briefly summarize the contributions of this work as follows. The number of Floquet bands in these models may be tuned by one's choice of experimental parameters. When an odd number of bands are present, the bands' Chern numbers \cite{HoGongPRL2012,KHMvsORDKR,ZhaoKickedHall} successfully predict the net number of chiral modes within each quasienergy gap according to the usual bulk-boundary correspondence rule \cite{RudnerPRX} (see Sec. \ref{3band-section} for details). However, there also occur in both models anomalous counter-propagating (ACP) modes\cite{ZhaoKickedHall} whose existence cannot be predicted from the Chern numbers. These ACP modes were first discovered in Ref. \cite{ZhaoKickedHall} in the lattice KHM under a particular choice of boundary condition (BC). Our work here builds on their finding in two ways. Firstly, we identify the existence of these modes in several other instances- namely under different BCs and also in a totally separate model in the form of the ORDKR lattice. Secondly, we show that these modes are in fact a weak topological effect and show results suggesting that they are related to the particular form of chiral symmetry (CS) operator describing the system. Moving on to consider an even number of bands, we report the existence of topologically protected $0$ and $\pi$ quasienergy modes in the ORDKR model but not in the KHM. These modes only occur along open BCs along one dimension but not along the other dimension.  We show that they are in fact governed by 1D topological invariants \cite{AsbothBBC,AsbothBBC1DSys}, demonstrating that they are once again a manifestation of a weak topological effect. The existence of these 1D invariants is again tied back to the particular form of CS operator. Our numerics also reveal that as certain parameters of the ORDKR model are tuned up, a large number of these topological edge modes occurs, together with a proliferation of Dirac cones in the quasienergy spectrum. This finding may be useful for quantum information processing with Floquet Majorana modes\cite{BarangerPRL} and the study of Dirac cones\cite{EsslingerNature2012,MontambauxPhysicaB,HasegawaPRB,SticletLongHoppingPRB2013,HouPRB2014,Tashima2014}. So far as we know, this paper is the first to consider weak topological effects in driven systems.

The outline of this paper is as follows. In Sec.~II, we introduce briefly the KHM and the ORDKR models, originally discussed in the quantum chaos literature and typically addressed in the angular momentum representation.  To study their edge state behaviour, we introduce 2D lattice versions of these two models, which we refer to as the Kicked Harper Lattice (KHL) and the Double Kicked Lattice (DKL) models, respectively. These lattice versions are mathematically identical to the original KHM and ORDKR (due to the equivalence between the lattice sites representation and the angular momentum representation) but are physically more meaningful. To lay the groundwork for later sections, we analyse the symmetries of the two models on a general level in Sec.~III. In Secs. IV and V, we specialize to 3-band and 2-band cases respectively in both models and study their weak topological edge state behaviour along different boundaries. The 3-band and 2-band cases are typical examples of odd-band and even-band behaviour in our driven systems. We point out the relationships between these topological states and the symmetries as well as bulk topological invariants present in the models. We conclude in Sec.~VI.

\section{Two dynamical models and their implementations on a lattice}

For completeness, we give a brief introduction to the background of the KHM and the ORDKR models which are both well-studied in the context of quantum chaos. We note in passing that a very recent study \cite{PlanckDrivenHallEffect} has argued that topological phenomena may emerge as a result of chaos, thus suggesting that the two seemingly disparate topics of chaos and topological insulators may even be related in a fundamental way.

The KHM displays chaos (given a suitable choice of parameters) when treated classically, yet its quantum version is simple enough for accurate numerical study. Insights on many topics have been gained from studies of this model. Such topics include metal-insulator transitions \cite{ABCinKHM,PhaseDiagKHM,DimerDecimation} and quantum eigenstate topology \cite{Leboeuf,danapre}. Remarkably, the KHM displays an unusual fractal-like quasienergy spectrum due to its close connection \cite{MetamorphosisPRL} with the famous Hofstadter butterfly spectrum \cite{Hofstadter}.

The ORDKR model is a particular example of modulated kicked rotors \cite{MRotors, Monteiro} and is another classically chaotic model which has yielded interesting features quantum mechanically. It displays intriguing features such as ratchet acceleration \cite{WangPRE2008} and exponential quantum spreading \cite{exponential}. This model has close connections with the KHM. Under an appropriate choice of parameters, the ORDKR also displays a Hofstadter-like quasienergy spectrum analogous to that of the KHM while at the same time displaying qualitatively different dynamics \cite{Gong2008proposalPRA,JMO}. Subsequent work \cite{WayneJMP} found that the spectra of the two models are identical provided that either an effective Planck constant parameter is irrational or a union of spectra over an added phase shift parameter (as we shall see later, this phase shift parameter may be regarded as the crystal momentum along the second dimension of a 2D model) is taken in both models.

The vast literature on KHM and our earlier studies of the ORDKR are based on the angular momentum representation, with both models displaying continuous Floquet bands due to a translational invariance in the angular momentum space.  This is not appropriate for the investigation of topological edge states because it is not clear how to introduce a physical boundary in the angular momentum space. For that reason we consider instead lattice versions of ORDKR and KHM.

Originally both KHM and ORDKR were 1D dynamical models. However, to study weak topological effects, we shall investigate 2D generalized versions of these two models. In particular, we start with a 2D square lattice of $L_{x} \times L_{y}$ sites, with both open and periodic boundary conditions along $x$ and $y$. We denote by $\ket{n_{x(y)}}$ the discrete lattice sites along $x \hspace{1mm} (y)$, where $n_{x(y)}=0, \cdots, L_{x(y)}-1$ \cite{lattice-rotor}. An arbitrary state in the Hilbert space is then written as $\ket{\psi} = \sum_{n_{x},n_{y}} \psi_{n_{x},n_{y}}\ket{n_{x}}\ket{n_{y}}$.
\subsubsection{DKL as a lattice version of ORDKR}
The first model we consider is a double-kicked lattice (DKL) model \cite{GongModRotorPRE2007, bose}, a lattice version of ORDKR, described by the following Hamiltonian:
\begin{equation}
H_{\text{DKL}} (t) = V(t) \hat{n}_{x}^{2} + \frac{1}{2}\sum_{n_{x}=0}^{L_{x}-1} \left( \hat{J}(t) \ket{n_{x}+1}\bra{n_{x}} + \hat{J}^{\dagger}(t) \ket{n_{x}}\bra{n_{x}+1} \right), \label{LDR2D}
\end{equation}
with
\begin{eqnarray}
V(t)=0; & \hspace{1cm} \hat{J}(t)=J_{1} \sum\limits_{n_{y} = 0}^{L_{y}-1} \ket{n_{y}+1}\bra{n_{y}} &  \hspace{0.5cm} \text{for} \hspace{0.3cm} 4m \leq t < 4m+1, \nonumber \\
V(t)=V; & \hat{J}(t)=0 & \hspace{0.5cm} \text{for} \hspace{0.3cm} 4m+1 \leq t < 4m+2 , \nonumber \\
V(t)=0; & \hat{J}(t)=J_{2} & \hspace{0.5cm} \text{for}  \hspace{0.3cm} 4m+2 \leq t < 4m+3, \nonumber \\
V(t)=-V; & \hat{J}(t)=0 & \hspace{0.5cm} \text{for}  \hspace{0.3cm} 4m+3 \leq t < 4(m+1), \label{ORDKR-periods-2D}
\end{eqnarray}
where $m \in \mathds{Z}$. The above Hamiltonian describes a time-periodic protocol consisting of four stages per period. During the first stage, the Hamiltonian describes a particle undergoing hopping in a diagonal fashion on the lattice. In the second stage, the particle is subject to a potential of strength $V$ which is quadratic along $x$ and independent of $y$. Next, the particle experiences a nearest-neighbour hopping of strength $J_{2}/2$ along only the $x$-direction. Finally, the particle experiences again the same potential that is quadratic along $x$, except with negative strength $-V$, meaning that this parabolic potential is inverted relative to the earlier one.  The consideration of a finite lattice (i.e., open boundary conditions) will reveal edge state properties, whereas applying periodic boundary conditions will reveal the bulk spectrum.
For the latter purpose, we may obtain a compact form of the Floquet operator for the DKL by introducing the translationally invariant crystal momentum states, defined by
\begin{equation}
\ket{k_{x(y)}} = \frac{1}{\sqrt{L_{x(y)}}} \sum_{n_{x(y)}=0}^{L_{x(y)}-1} \ket{n_{x(y)}} e^{-ik_{x(y)}n_{x(y)}} \label{fourier-trans}
\end{equation}
along $x(y)$, where $k_{x(y)}=-\pi + j \times 2\pi/L_{x(y)} $ and $j = 0 , 1, \cdots, L_{x(y)}-1 $. Using the above equation, one may show that the Floquet operator which propagates from $t=0$ to $t=4$ takes the form
\begin{equation}
U_{\text{DKL}} (J_{2},V,J_{1}) = e^{i \hat{n}_{x}^{2} V}e^{-iJ_{2}\cos(\hat{k}_{x})}e^{-i \hat{n}_{x}^{2} V}e^{-iJ_{1}\cos(\hat{k}_{x}+\hat{k}_{y})}, \label{lattice-UORDKR-ky}
\end{equation}
where we have chosen to work in dimensionless units such that $\hbar=1$.
Within each $k_{y}$ subspace, the above $U_{\text{DKL}}$ is seen to be of precisely the same form as the Floquet operator of a 1D ORDKR treated in the angular momentum space \cite{Gong2008proposalPRA}, where $\hat{n}_x$ plays the role of the angular momentum operator, $\hat{k}_x$ the role of an angular variable, and $k_{y}$ the role of a phase shift parameter.  A similar perspective was discussed by others \cite{KrausPumpingPRL} where the 2D model was referred to  as the ``ancestor" of 1D models within each $k_{y}$ subspace.

Throughout, we denote the quasienergy and the associated eigenstate of a Floquet operator $U$ as $\omega_{n}$ and $\ket{\psi_{n}}$ respectively \cite{Quasienergies}, with $\hat{U} \ket{\psi_{n}} = e^{-i \omega_n} \ket{\psi_{n}}$. Since the quasienergy is only defined modulo $2\pi$, we define the quasienergy Brillouin zone (BZ) as ranging from $-\pi$ to $\pi$.
By choosing $V$ such that $V= \pi M/N$, where $M,N \in \mathds{Z}$, the Floquet operator $U_{\text{DKL}}$ becomes periodic in the $\ket{n_{x}}$ representation with period $N$. Bloch's theorem then yields that we will have a quasienergy spectrum consisting of $N$ bands. For low values of $J_{1,2}$, the spectrum consists of $N$ bands separated by large gaps. For a fixed value of $V$, increasing the values of $J_{1,2}$ will cause the quasienergy bands to broaden and occupy more space within the quasienergy BZ. As $J_{1,2}$ increase beyond certain special values, the quasienergy bands will touch and re-separate, possibly causing a topological phase transition. Later, we shall study the spectra obtained as $J_{1,2}$ increase for different $ V= \pi M/N$ and observe the effects that the topological phase transitions have on the topological invariants and related edge states.

\subsubsection{KHL as a lattice version of KHM}
Here we consider a lattice version of KHM, which we refer to as the kicked Harper lattice (KHL), described by Hamiltonian
\begin{eqnarray}
H_{\text{KHL}}(J,R,b)&=& \frac{J}{2} \sum_{n_{x} =0}^{L_{x}-1}\left(\ket{n_{x}+1}\bra{n_{x}}+\ket{n_{x}}\bra{n_{x}+1}\right) + \nonumber \\
&& \frac{R}{2} \sum_{n_{x},n_{y}=0}^{L_{x}-1,L_{y}-1} \left( e^{in_{x}b}\ket{n_{x},n_{y}}\bra{n_{x},n_{y}+1} +\text{h.c} \right) \sum_{m} \delta (t-m) \nonumber \\
&=& J \cos (\hat{k}_{x}) +  R \cos(\hat{n}_{x}b  - \hat{k}_{y})\sum_{m} \delta (t-m), \label{LKHM-2D}
\end{eqnarray}
with $m \in \mathds{Z}$, where we have made use of Eq. (\ref{fourier-trans}) in order to obtain the second line that applies to the case under periodic boundary condition (for the purpose of understanding the bulk spectrum). The above Hamiltonian is directly related to a solid-state system subject to a kicking control field and in Ref.~\cite{ZhaoKickedHall} it was called a kicked Hall system. The Floquet operator evolving states from time $t=0^{+}$ to time $t=1^{+}$ is then given by
\begin{equation}
U_{\text{KHL}}(J,R,b)=e^{-iR\cos(\hat{n}_{x} b -\hat{k}_{y})} e^{-iJ\cos(\hat{k}_{x})}.  \label{lattice-UKHM-2D}
\end{equation}
Within each single $k_{y}$ subspace, this is indeed the familiar form of the 1D KHM Floquet operator, with $k_{y}$ playing the role of a phase shift parameter as introduced in our early studies \cite{WangPRE2008,WayneJMP,KHMvsORDKR}. By choosing $b = 2\pi M /N$, where $M,N \in \mathds{Z}$, we again obtain an $N$-band quasienergy spectrum just like we did for the DKL Floquet operator. This completes our construction of the lattice versions of ORDKR and KHM.

\section {General Analysis of Symmetries} \label{GASymmetries}
\subsection{Brief Review on Chiral Symmetry in Driven Systems}
The work of Ref.~\cite{KitagawaTopoChar} suggested that given a Floquet operator $\hat{U}$, assuming that there is no winding of quasienergy across the BZ seen in the spectrum (i.e., the quasienergy spectrum remains between $-\pi$ and $\pi$ at all points in the BZ), one may extract an effective static Hamiltonian, $\hat{H}_{\text{eff}}$, via the relation
\begin{equation}
\hat{U} \equiv e^{-i \hat{H}_{\text{eff}}}, \label{Heff-defn}
\end{equation}
and classify $\hat{H}_{\text{eff}}$ according to the tenfold classification scheme for static systems \cite{Tenfoldway}, thus effectively classifying $\hat{U}$. Following this approach, a Floquet operator $\hat{U}$ is said to possesses chiral symmetry (CS) \cite{KitagawaTopoChar,AsbothBBC,Tenfoldway} if there exists a unitary and Hermitian operator $\Gamma$ such that
\begin{equation}
\Gamma \hat{U} \Gamma^{\dagger} = \hat{U}^{-1}, \label{CS-condtn}
\end{equation}
with $\Gamma$ obeying $\Gamma^{2}=\mathds{1}$ \cite{KitagawaTopoChar, AsbothBBC}. We shall refer to $\Gamma$ as the CS operator. An ambiguity naturally arises at this point. Namely, there is an arbitrary choice of which one-period time interval to choose for a Floquet operator to propagate across. It turns out that different choices can lead to $\hat{H}_{\text{eff}}$ possessing different symmetries or none at all. We follow the strategy introduced in Ref.~\cite{AsbothBBC} and seek ``symmetric time frames", which are defined as choices of time frames resulting in Floquet operators $\hat{U}$ of the form
\begin{equation}
\hat{U} = \hat{F}\hat{G},
 \end{equation}
 where $\hat{F}$ and $\hat{G}$ are unitary operators related with each other via the CS operator:
 \begin{equation}
 \Gamma \hat{F} \Gamma = \hat{G} ^{-1}.
 \end{equation}
It is trivial to prove that once this relation is obeyed, so too is the CS condition in Eq. (\ref{CS-condtn}). Such symmetric time frames do not exist for arbitrary Floquet operators but do in the case of the DKL and KHL, as we shall prove shortly. It is easy to see that if a symmetric time frame exists corresponding to a Floquet operator $\hat{U}' = \hat{F}\hat{G}$ exists, then there must also be a second symmetric time frame corresponding to Floquet operator $\hat{U}'' = \hat{G}\hat{F}$ \cite{AsbothBBC}. We note for general interest that Floquet operators possessing CS in symmetric time frames in general do \emph{not} obey CS in arbitrary (non-symmetric) time frames. This fact hints at the existence of some generalized form of chiral symmetry which is present regardless of the choice of time frame. If such a generalization exists, it has yet to be found, but we do not tackle this issue in the present work.

\subsection{Symmetry operators for DKL and KHL}
The Floquet operator for the DKL model in a symmetric time frame from $t=2.5$ to $t=6.5$ (cf. Eqs. (\ref{LDR2D}) and (\ref{ORDKR-periods-2D}) ) reads
\begin{equation}
U'_{\text{DKL}} (J_{2},V,J_{1}) = e^{-i\frac{J_{2}}{2}\cos(\hat{k}_{x})}e^{-i \hat{n}_{x}^{2} V}e^{-iJ_{1}\cos(\hat{k}_{x}+\hat{k}_{y})}e^{i \hat{n}_{x}^{2} V}e^{-i\frac{J_{2}}{2}\cos(\hat{k}_{x})}, \label{lattice-UORDKR-2D-Symm-frame}
\end{equation}
where
\begin{eqnarray}
\hat{F} & \equiv e^{-i J_{2}\cos(\hat{k}_{x})/2}
e^{-i\hat{n}_{x}^{2}V}e^{-iJ_{1}\cos(\hat{k}_{x}+\hat{k}_{y})/2} \nonumber \\
\hat{G} & \equiv  e^{-iJ_{1}\cos(\hat{k}_{x}+\hat{k}_{y})/2}e^{i \hat{n}_{x}^{2}V}e^{-i J_{2}\cos(\hat{k}_{x})/2}.
\end{eqnarray}
The CS operator is given by
\begin{equation}
\Gamma_{\text{DK}} = e^{i \hat{n}_{x} \pi}. \label{CS-OP-DKL}
\end{equation}
It is clear that
\begin{equation}
\Gamma_{\text{DK}} ^{2} = 1
 \end{equation}
 and
 \begin{equation}
 \Gamma_{\text{DK}} = \Gamma_{\text{DK}}^{\dagger} = \Gamma_{\text{DK}}^{-1}
  \end{equation}
  since $\hat{n}_{x}$ has only integer eigenvalues. Making use of the fact that
\begin{equation}
e^{i \hat{n}_{x} \pi} f(\hat{k}_{x}) e^{-i \hat{n}_{x} \pi} = f(\hat{k}_{x}+\pi) \label{translatn-in-k}
\end{equation}
for an arbitrary function $f$, we see that
\begin{eqnarray}
\Gamma_{\text{DK}} U'_{\text{DKL}} \Gamma_{\text{DK}} &=& \Gamma_{\text{DK}} \hat{F} \hat{G} \Gamma_{\text{DK}} \nonumber \\
&=& \Gamma_{\text{DK}} \hat{F} \Gamma_{\text{DK}}^{2} \hat{G} \Gamma_{\text{DK}} \nonumber \\
&=&   \hat{G} ^{-1}  \hat{F} ^{-1} \nonumber \\
&=& U_{\text{DKL}}^{'-1}. \label{CS-DKL}
\end{eqnarray}
This proves that the DKL Floquet operator possesses CS in a symmetric time frame.

Next, we analyse the symmetry of the KHL Floquet operator. Defining the Floquet operator as propagating states across the symmetric time frame from $t=0.5$ to $t=1.5$ (cf. Eq. (\ref{LKHM-2D})), we obtain
\begin{equation}
U'_{\text{KHL}}(J,R,b)=e^{-i\frac{J}{2}\cos(\hat{k}_{x})}e^{-iR\cos(\hat{n}_{x} b -\hat{k}_{y})} e^{-i\frac{J}{2}\cos(\hat{k}_{x})}.  \label{lattice-UKHM-2D-symm-frame}
\end{equation}
The CS operator of the above model is given by
\begin{equation}
\Gamma_{\text{KH}} = e^{i\hat{n}_{x}\pi} e^{i\hat{n}_{y}\pi}. \label{CS-OP-KHL}
\end{equation}
 Clearly,
 \begin{equation}
 \Gamma_{\text{KH}}^{2} = \mathds{1}
  \end{equation}
  and
  \begin{equation}
  \Gamma_{\text{KH}} = \Gamma_{\text{KH}}^{-1} = \Gamma_{\text{KH}}^{\dagger}.
   \end{equation}
   The CS condition may be easily verified using Eq. (\ref{translatn-in-k}) as follows:
\begin{eqnarray}
& & \Gamma_{\text{KH}} e^{-i\frac{J}{2}\cos(\hat{k}_{x})}e^{-iR\cos(\hat{n}_{x} b -\hat{k}_{y})} e^{-i\frac{J}{2}\cos(\hat{k}_{x})} \Gamma_{\text{KH}} \nonumber \\
&=& e^{-i\frac{J}{2}\cos(\hat{k}_{x}+\pi)}e^{-iR\cos(\hat{n}_{x} b -(\hat{k}_{y}+\pi))} e^{-i\frac{J}{2}\cos(\hat{k}_{x}+\pi)} \nonumber \\
&=& e^{i\frac{J}{2}\cos(\hat{k}_{x})}e^{iR\cos(\hat{n}_{x} b -\hat{k}_{y})} e^{i\frac{J}{2}\cos(\hat{k}_{x})} \nonumber \\
&=& U_{\text{KHL}}^{'-1}(J,R,b). \label{CS-2D-LKH}
\end{eqnarray}
Thus, the KHL model possesses CS. We note that it may be shown that both models still obey the CS condition of Eq. (\ref{CS-condtn}) with their respective CS operators even when open BCs are taken along one or both axes.

In the following sections, we shall study the two models when $V=\pi/N$ and $b=2\pi/N$, where the $N=3$ and $N=2$ cases will be considered in Secs. IV and V respectively. It is useful to set up some notation here for this purpose. Later, we will consider the two models under both periodic and open boundary conditions (BCs) in order to study bulk-boundary correspondence. When periodic BCs are taken along $x$ and $y$, we shall write the Floquet operators of both models, referred to generically as $U$, in the crystal momentum representation which reflects their translational invariance. This representation is defined as follows. The lattice sites along $x$, $\{ \ket{n_{x}} \}$, are divided into sublattices labelled $P$, where $P = 0 ,\cdots, N-1$, each of size $S \equiv L_{x}/N$. The sites of sublattice $P$ are denoted as $\ket{\bar{n}_{x},P} \equiv \ket{n_{x}=P+\bar{n}_{x}N}$, where $\bar{n}_{x}=0, \cdots , S-1$. We then define reciprocal lattice (crystal momentum) states of $\ket{\bar{n}_{x},P}$ via the Discrete Fourier Transform as
\begin{equation}
\ket{\bar{k}_{x},P} = \frac{1}{\sqrt{S}}\sum _{\bar{n}_{x}} \ket{\bar{n}_{x},P} e^{-i \bar{n}_{x} \bar{k}_{x}}, \label{Discrete-FT}
\end{equation}
where $ \bar{k}_{x} = -\pi, -\pi + 2\pi/S, \cdots, \pi - 2\pi/S$. Note that the reciprocal lattice states $\ket{\bar{k}_{x},P}$ are Bloch-periodic (i.e., periodic up to a phase factor) in the lattice space over every $N$ sites, unlike the $\ket{k_{x}}$ seen earlier which are Bloch-periodic over every 1 site. Along the $y$ direction, we simply work in the representation of $\ket{k_{y}}$ defined in Eq. (\ref{fourier-trans}). In this representation, our Floquet operators will take the form
\begin{equation}
\hat{U} = \sum_{\bar{k}_{x},k_{y}} [ U(\bar{k}_{x},k_{y})] \otimes \ket{\bar{k}_{x}}\bra{\bar{k}_{x}} \otimes \ket{k_{y}}\bra{k_{y}}, \label{standard-form}
\end{equation}
where $[ U(\bar{k}_{x},k_{y})]$ is an $N \times N$ unitary matrix describing the coupling between the $P$ (sublattice) degrees of freedom within each $\bar{k}_{x}$ space. A recurring theme in our analysis in the following sections will be to analyse the effect of the CS operators in Eqs. (\ref{CS-OP-DKL}) and (\ref{CS-OP-KHL}) on the Floquet matrices $[ U(\bar{k}_{x},k_{y})]$ and how they transform these matrices into their inverses within the same or different $(\bar{k}_{x},k_{y})$ subspace. When we take open BCs along one direction and periodic BCs along the other, the above decomposition into different momentum spaces will only be possible along one direction and the Floquet operator will take the form
\begin{equation}
\hat{U} = \sum_{k} [U(k)] \otimes \ket{k}\bra{k}, \label{standard-form2}
\end{equation}
where $k$ here may refer to $\bar{k}_{x}$ or $k_{y}$ depending on the direction along which periodic BCs are taken. The matrix $[U(k)]$ then describes the coupling within each $k$-space. This notation will be useful for discussing the topological behaviour of our models for the $3$-band and $2$-band cases.

Before ending this section, we elaborate upon the 1-1 mapping between the two models that we alluded to in the introduction. In Ref. \cite{KHMvsORDKR}, we proved that when $V = b/2 = \pi M/N$ for all odd $N$, the matrices $[ U_{\text{DKL}}(\bar{k}_{x},k_{y})]$ and $[ U_{\text{KHL}}(\bar{k}_{x}+ N\pi,k_{y}-\bar{k}_{x}/N)] $ are related to each other by a unitary transformation whenever $J_{1}=R$ and $J_{2}=J$. We have since discovered that an analogous mapping also holds for all even $N$.\cite{mappingnote} This exact mapping result means that the eigenstates of the two $N\times N$ matrices (i.e., the full Floquet operator's eigenstates projected onto one unit cell) are related to each other by unitary transformations. Since it is these reduced eigenstates that feature in expressions for topological invariants, by studying the topological properties of the two models, we are in fact studying what happens to a system's topological properties under a rearrangement of eigenvalues and eigenstates on the crystal momentum BZ. We shall see in the following sections that this rearrangement does not affect the system's strong topological properties (i.e., boundary shape-independent edge modes described by 2D invariants), but gives rise to differences which turn out to be weak topological properties (i.e., boundary shape-dependent edge modes described by 1D invariants).

\section{Topological States in 3-band cases} \label{3band-section}

In this section we study the effect of the CS in 3-band cases by setting $V=\pi/3$ in $U'_{\text{DKL}}$ and $b=2\pi/3$ in $U'_{\text{KHL}}$. We first report our numerical data from both models before analysing their symmetries in order to gain insight.

\subsection{Numerical Results and Discussions}
Taking periodic BCs along $x$ and $y$ and writing the Floquet operators in the standard form of Eq. (\ref{standard-form}) and numerically diagonalizing the $3\times 3$ matrices $[ U'_{\text{DKL}}(\bar{k}_{x},k_{y})]$ and $[ U'_{\text{KHL}}(\bar{k}_{x},k_{y})]$ across the entire $(\bar{k}_{x},k_{y})$ BZ, we obtain $3$ quasienergy bands for both models. These quasienergy bands are known to possess nonzero Chern numbers \cite{HoGongPRL2012,KHMvsORDKR,ZhaoKickedHall} defined by
\begin{equation}
C_{n} = \frac{i}{2\pi} \oint d{\bf k} \cdot \bra{\psi_{n}(\bar{k}_{x},k_{y})} \nabla_{{\bf k}} \ket{\psi_{n}(\bar{k}_{x},k_{y})},
\end{equation}
where we have denoted the eigenstates of $[ U'_{\text{DKL}}(\bar{k}_{x},k_{y})]$ and $[ U'_{\text{KHL}}(\bar{k}_{x},k_{y})]$ generically as $\ket{\psi_{n}(\bar{k}_{x},k_{y})}$, $n=1,2,3$ is the band index and ${\bf k} \equiv (\bar{k}_{x},k_{y})$. These Chern numbers will be used for the study of bulk-boundary correspondence in both models. The Chern numbers are defined only under periodic BCs and are thus referred to as bulk invariants, as opposed to the numbers of topological edge modes which are invariants defined under open BCs.

\begin{figure}
\begin{center}$
\begin{array}{cc}
\includegraphics[height=!,width=7.5cm] {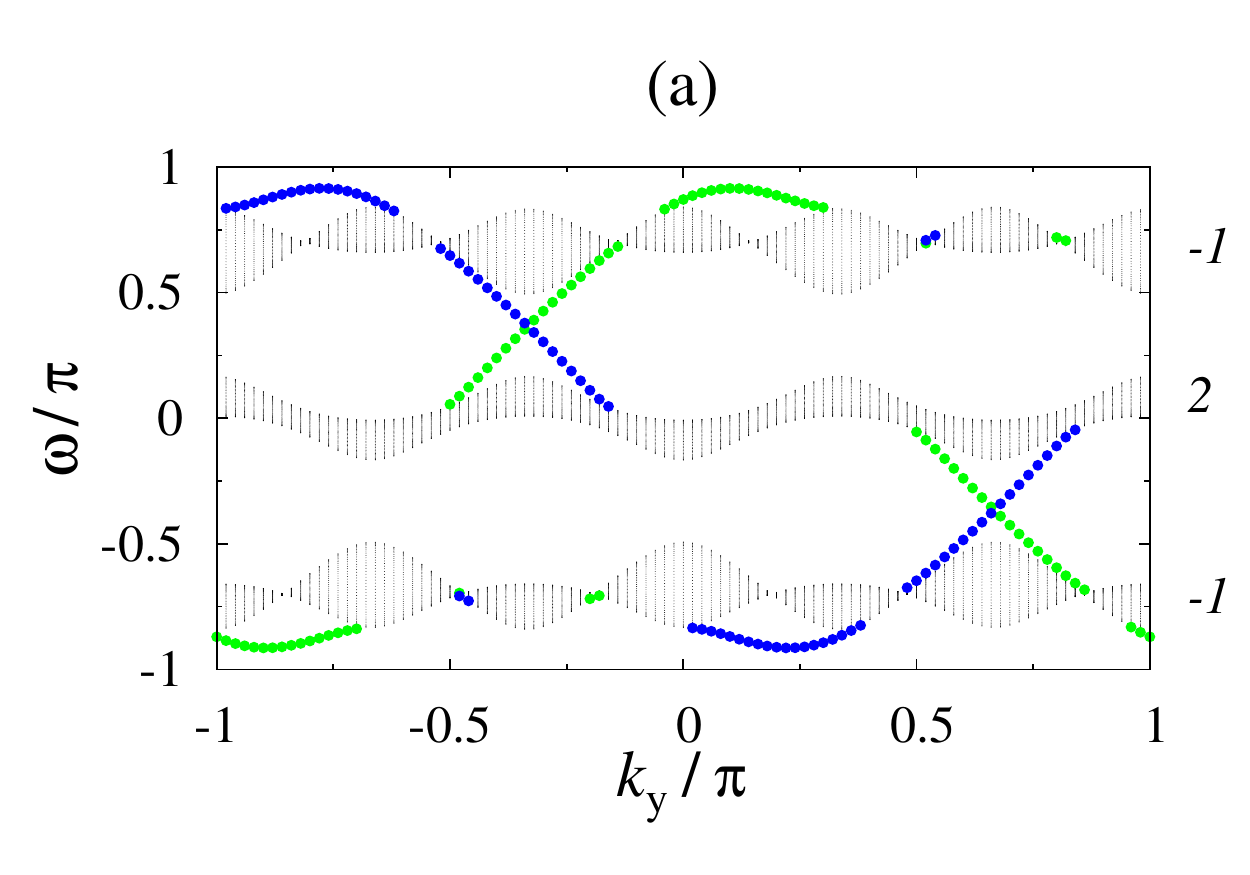}  &
\includegraphics[height=!,width=7.5cm] {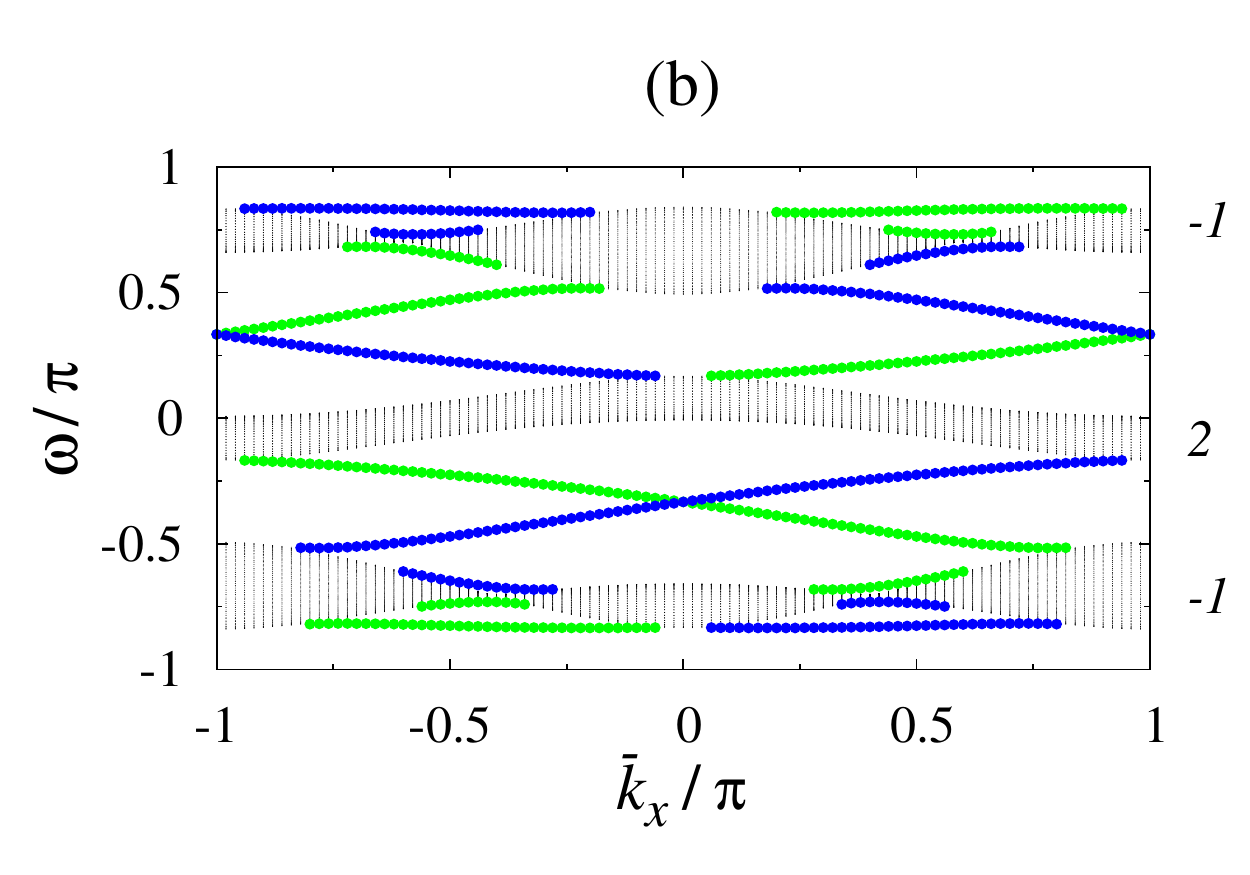} \\
\includegraphics[height=!,width=7.5cm] {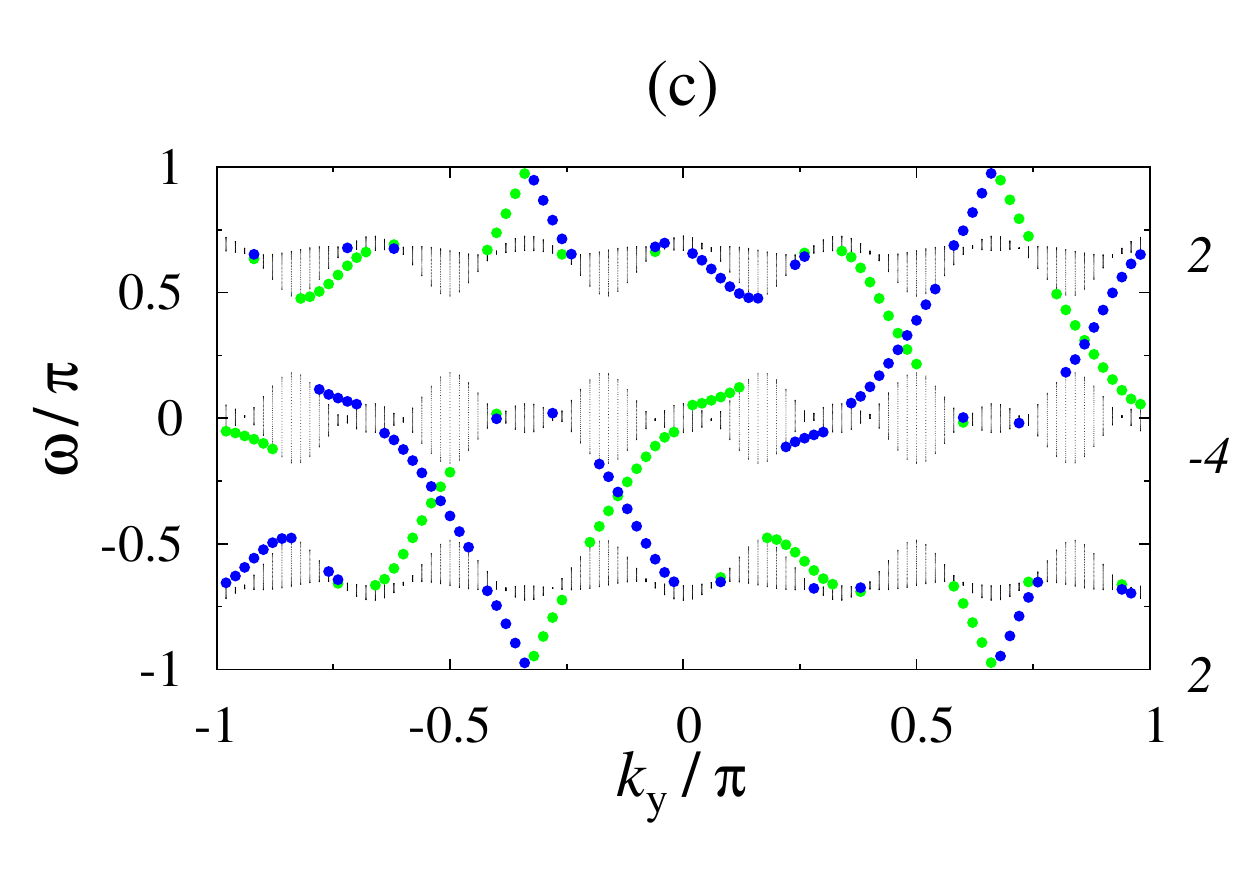} &
\includegraphics[height=!,width=7.5cm] {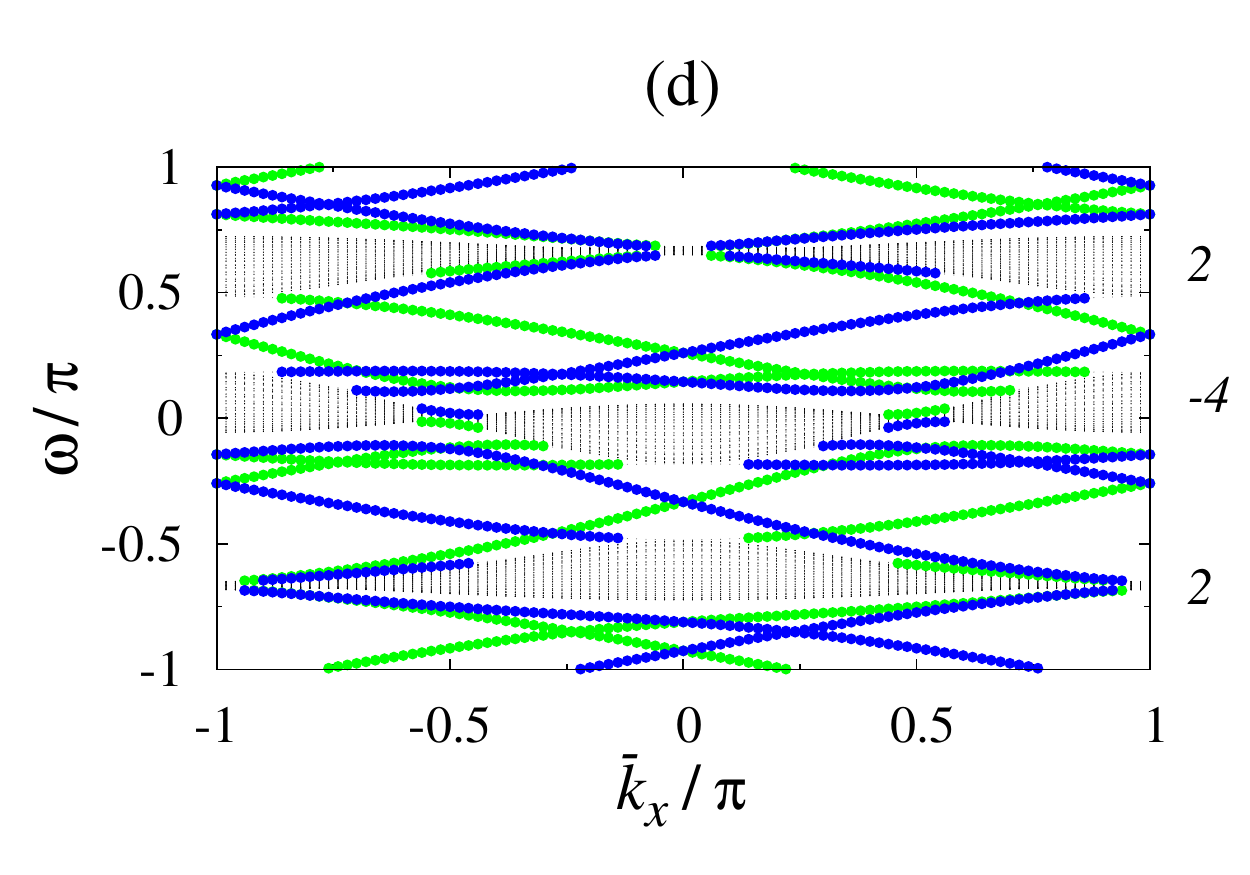} \\
\includegraphics[height=!,width=7.5cm] {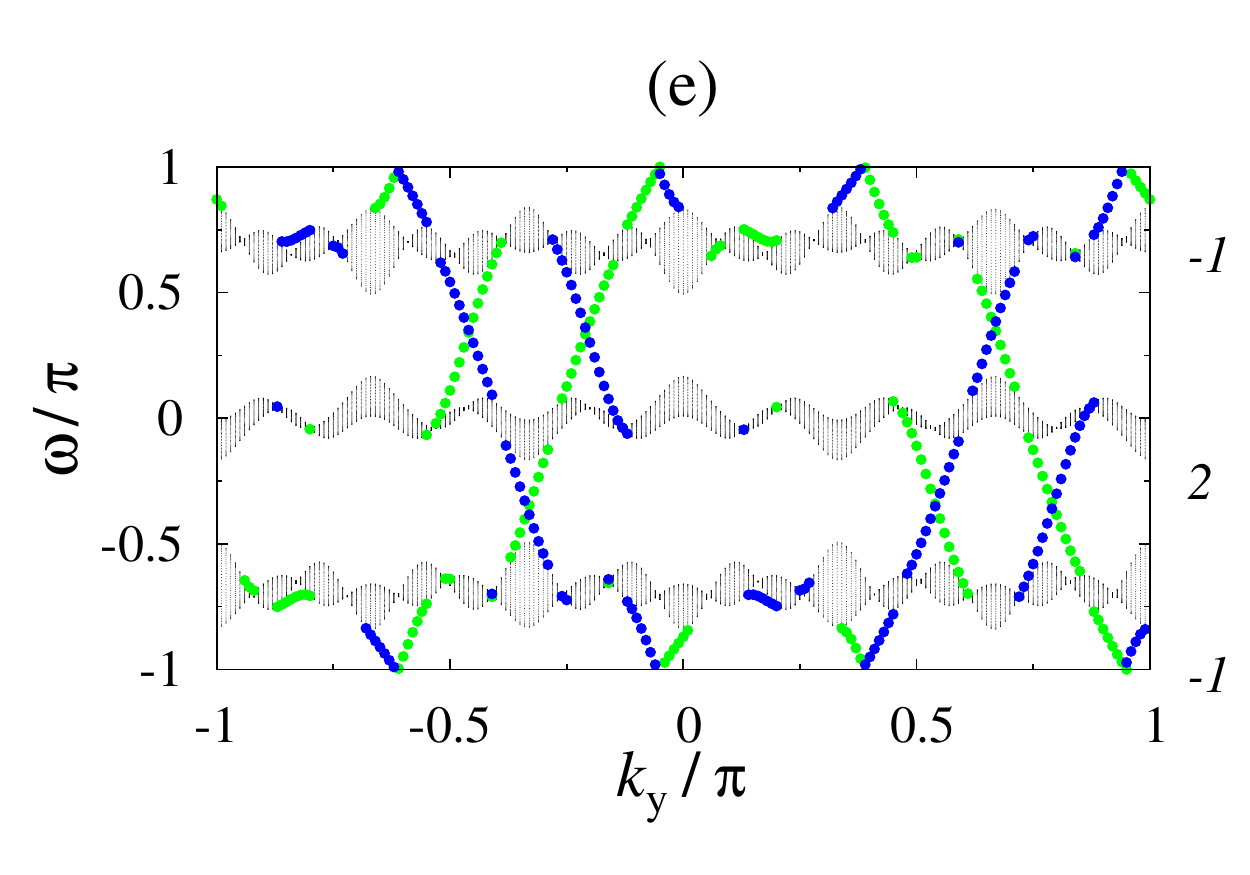} &
\includegraphics[height=!,width=7.5cm] {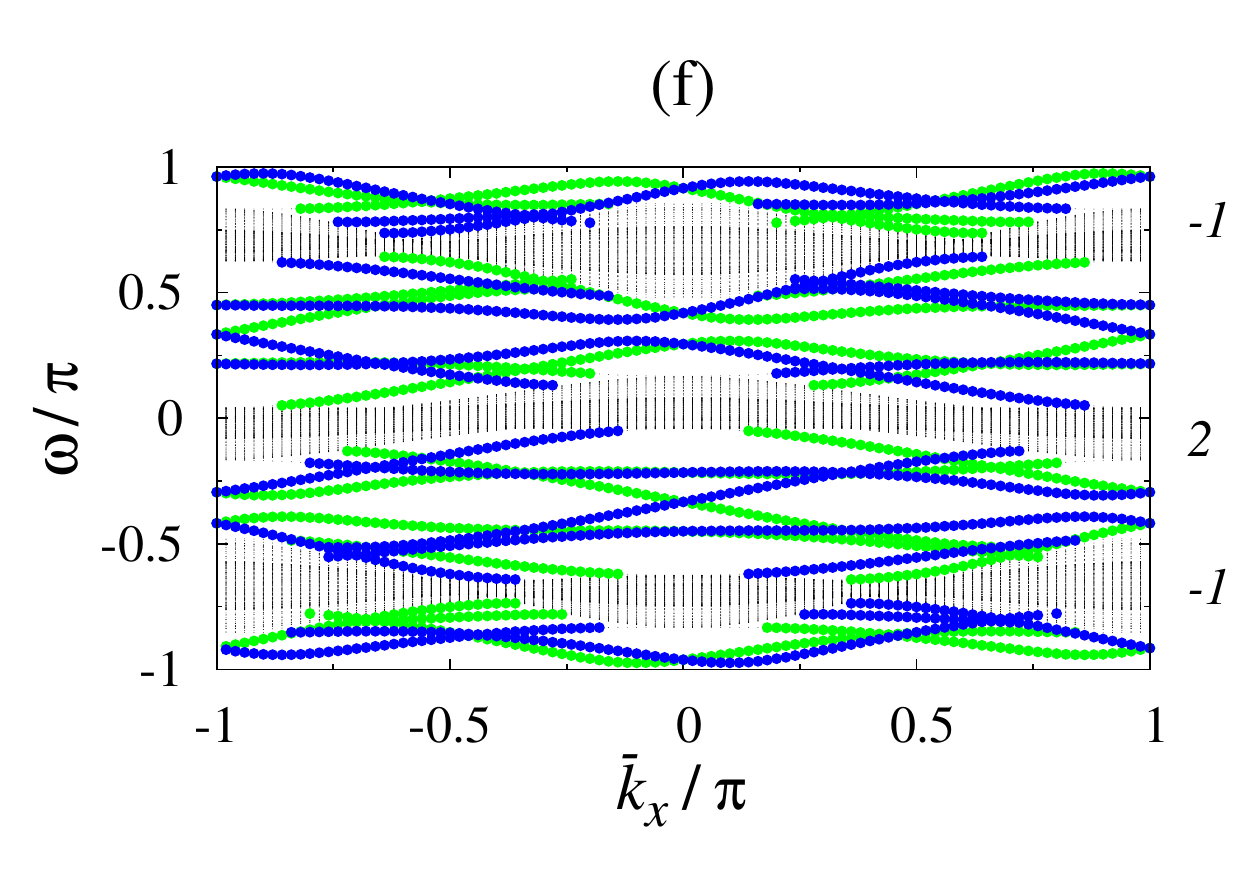}
\end{array}$
\end{center}
\caption{(color online). The QE spectra for the kicked Harper lattice model at $b=2\pi/3$, $J=2\pi /3$ under open BCs along $x$ ($y$) with $R=\pi,2\pi,3\pi$ are displayed in panels (a),(c),(e) [ (b), (d), (f) ] respectively. The Chern number bulk-boundary correspondence rule is obeyed regardless of choice of boundary. However, a comparison betwen (c) [(e)] and (d) [(f)] reveals that the number of ACP modes changes with boundary. Chern numbers of bulk bands are indicated on the right side of each figure panel.  Black (blue) lines in the spectrum gaps represent edges state on the left, whereas gray (green) lines in the spectrum gaps
  represent edge states on the right.  Here and in all other figures, variables are plotted in dimensionless units.} \label{khm-lattice-3band}
\end{figure}

The authors of Ref. \cite{ZhaoKickedHall} previously studied the KHL model (referred to there as a kicked quantum Hall system) under open BCs along $x$ (i.e., the edges are parallel to the $y$-axis).
As the starting point for our discussion, we reproduce some of their results in Figs. \ref{khm-lattice-3band}(a), (c) and (e)\cite{credit}.
As pointed out by these authors, under certain parameter choices (see Figs. \ref{khm-lattice-3band}(c) and (e)), ACP modes appear within certain gaps.
To be precise, ACP modes are defined as chiral modes on the same edge within the same quasienergy gap having opposite chirality.
These modes are of interest because when present, there is no known way of using bulk topological invariants to predict the actual number of chiral edge modes (see below for details). The ACP modes pointed out by Ref. \cite{ZhaoKickedHall} thus reveal a gap in the current understanding of bulk-boundary correspondence.
We note that similar ACP modes have also been numerically demonstrated in a static spinless system in a weak topological phase \cite{Hatsugai-2013JpnSoc}. For the usual strong static topological systems in class A (i.e., integer quantum Hall insulators), however, such ACP modes do not occur.

To make the significance of the ACP modes clear to a wide audience, let us briefly review here the current state of knowledge for systems with bands of non-zero Chern numbers. We note that by `systems', we refer to both static and driven systems whose Hamiltonians and effective Hamiltonians respectively are found in class A \cite{class-note} and by `bands', we refer to both energy and quasienergy bands.
It is known\cite{HatsuPRL1993,HatsuPRB1993} that in such systems, taking open BCs along one direction and keeping periodic BCs along the other (perpendicular) direction (i.e., a cylinder geometry), for any band $n$, the net chirality of edge modes (i.e., the total count of chiral modes signed according to their chirality) on each boundary in the gap above it subtracted by the same in the gap below it must be equal to the band's Chern number calculated under periodic BCs.
The Chern number of a band is thus the difference in net chirality of the edge states in the gaps above and below it.
In the static case (assuming no ACP modes are present), energies are bounded from below (i.e., no energies exist below the lowest bulk band) and given all the bands' Chern numbers, one may deduce the exact number of chiral edge states within each gap. In driven systems, due to the fact that quasienergy is only defined modulo $2\pi$, knowing all the bands' Chern numbers still leaves one unable to determine the number of edge modes, as pointed out in Ref. \cite{RudnerPRX}. In a dramatic example, these authors showed that a system whose bands all possess zero Chern number is still able to host topological chiral edge modes.
To remedy this ambiguity, they formulated a bulk winding number invariant that uniquely determines the net chirality of edge states in each gap.
In the absence of ACP modes, this winding number uniquely determines the number of chiral edge modes within each quasienergy gap.
However, when ACP modes \emph{are} present, the winding number still fails to tell us the actual number of chiral quasienergy modes present. At the time of writing, there is no known way to determine the actual number of modes from a bulk invariant whenever ACP modes are present. We note that in the static case, ACP modes have only been numerically observed (see Fig. 3 of Ref. \cite{Hatsugai-2013JpnSoc}) when the bulk bands all have Chern numbers equal to zero. In the kicked Hall system of Ref. \cite{ZhaoKickedHall} and in the DKL model (see below), however, the ACP modes occur even though the bands have nonzero Chern numbers. This suggests that the situation for driven systems may be quite different from that in static systems.

Moving beyond the numerical data of Ref. \cite{ZhaoKickedHall}, we compute the quasienergy spectra of the KHL model under open BCs along $y$ as shown in Figs. \ref{khm-lattice-3band}(b), (d) and (f). A comparison of Figs. \ref{khm-lattice-3band}(a), (c), (e) with (b), (d), (f) respectively then shows that while the Chern number bulk-boundary correspondence rule still holds true under open BCs along $y$, the ACP modes do not always persist. More specifically, the ACP modes are present along the open boundaries along $y$ when $J=2\pi/3,\hspace{1mm} R=2\pi$ (see Fig. \ref{khm-lattice-3band}(d)), but are absent when $R$ is increased to $3\pi$ (see Fig. \ref{khm-lattice-3band}(f)). This is in contrast to taking open BCs along $x$ which results in the  ACP appearing along the $x$-boundaries for both cases (see Figs. \ref{khm-lattice-3band}(c) and (e)). This dependence on boundary choice suggests that the ACP modes are a weak topological effect. Our finding here is similar to that in Ref. \cite{Hatsugai-2013JpnSoc} of weak topological ACP modes in a static 2D system, with the difference being that here we are dealing with a driven system.

\begin{figure}
\begin{center}$
\begin{array}{cc}
\includegraphics[height=!,width=7.5cm] {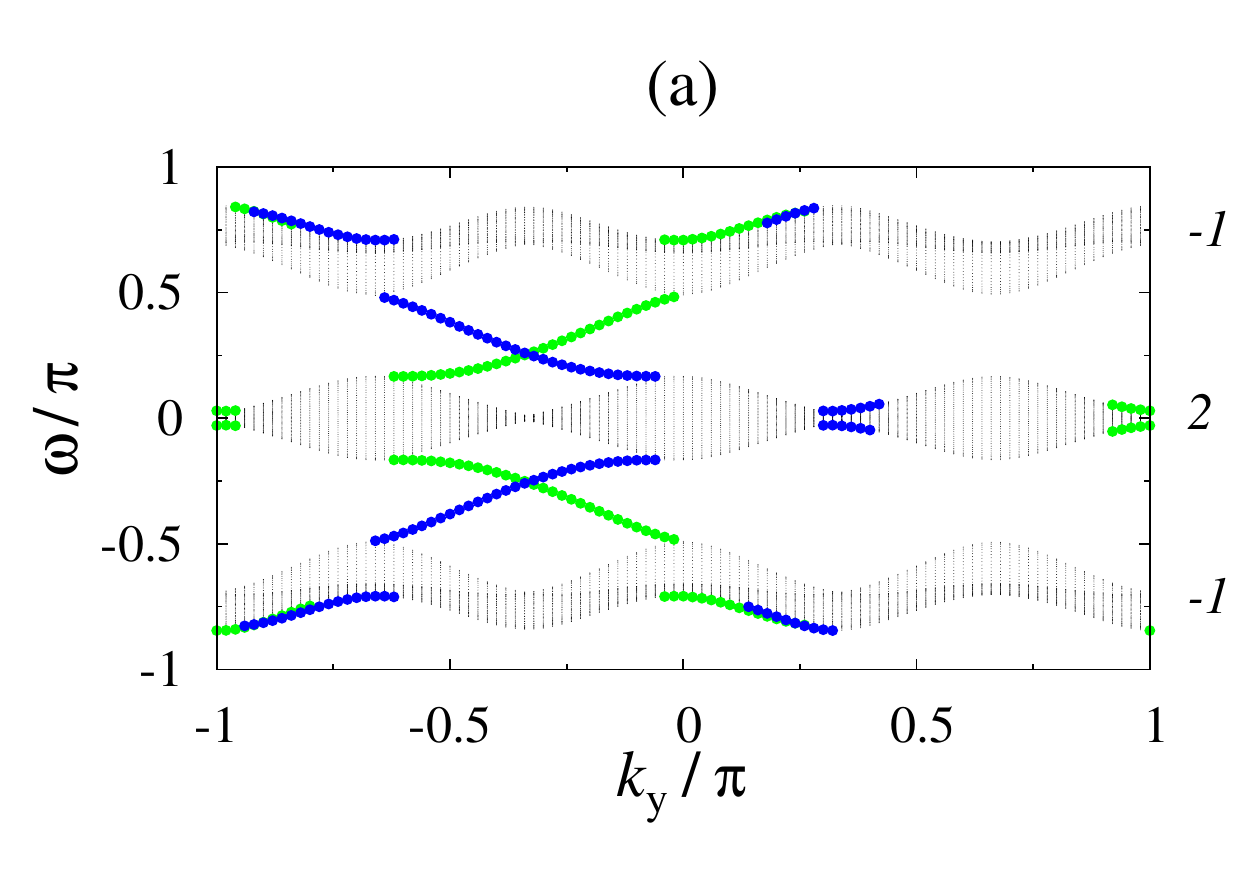}  &
\includegraphics[height=!,width=7.5cm] {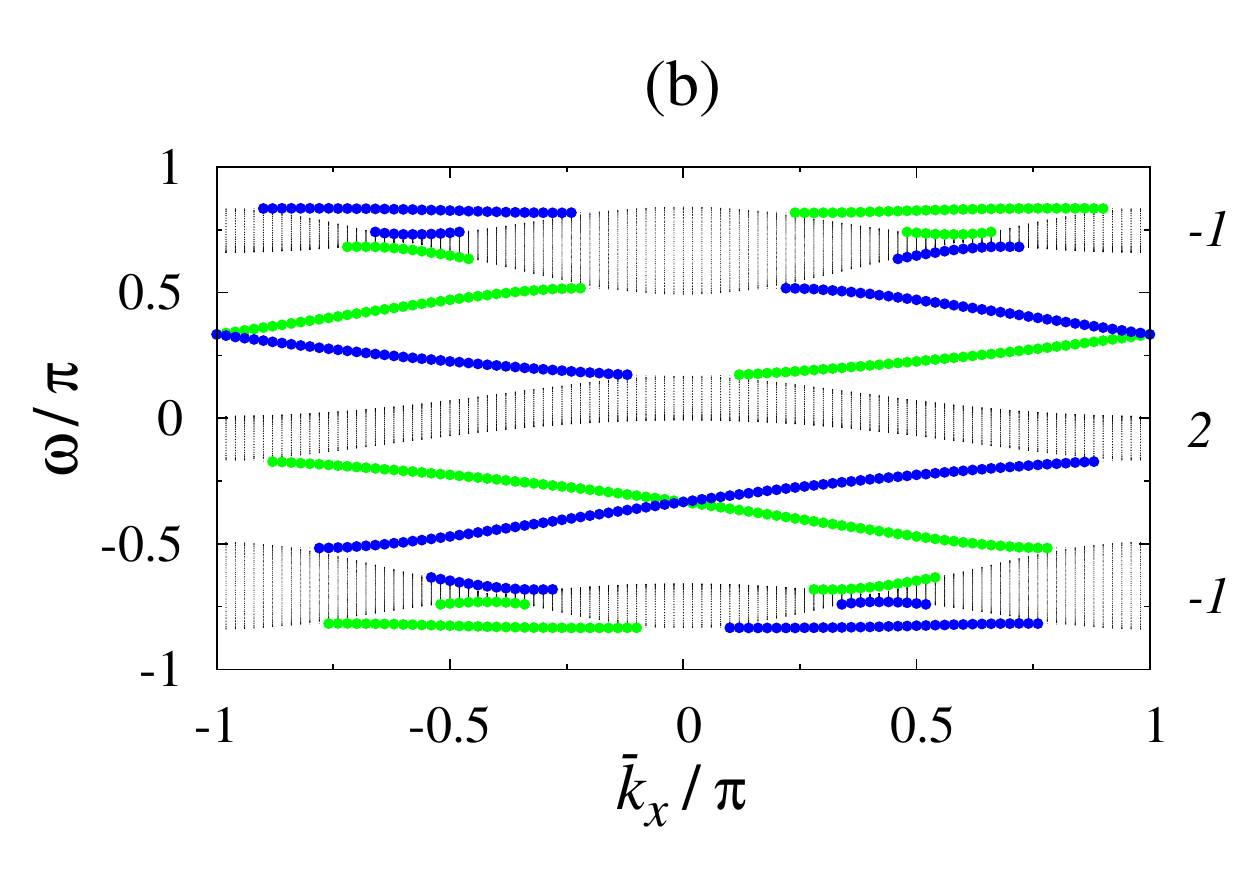} \\
\includegraphics[height=!,width=7.5cm] {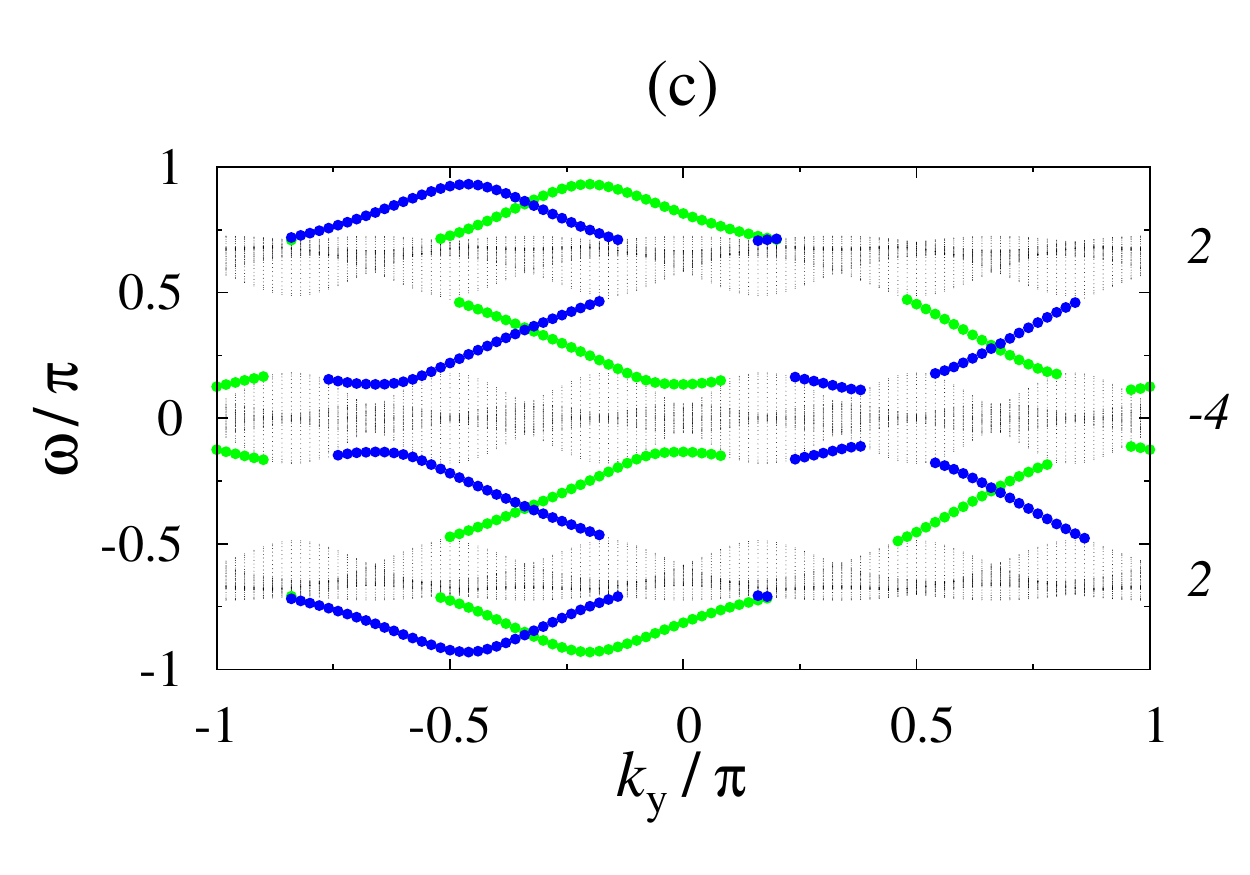} &
\includegraphics[height=!,width=7.5cm] {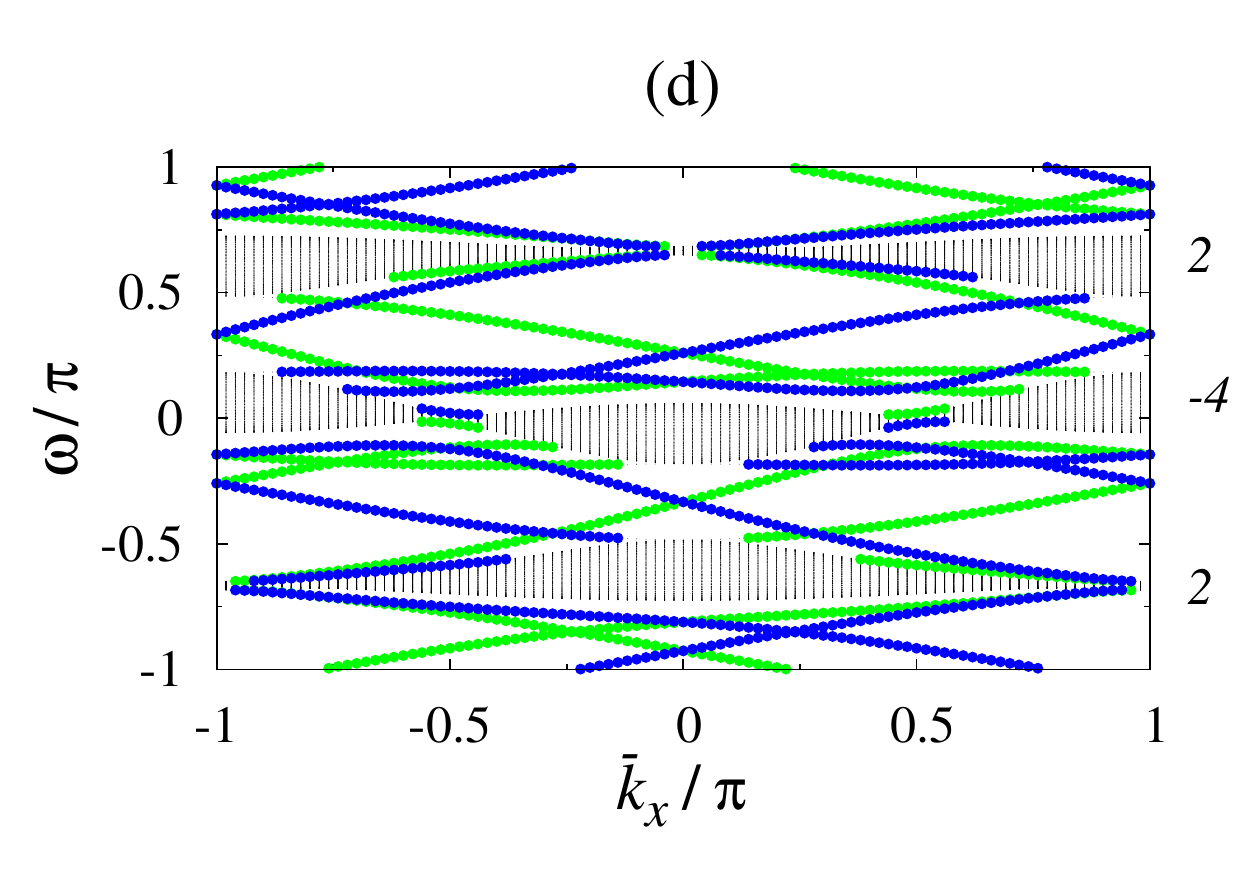} \\
\includegraphics[height=!,width=7.5cm] {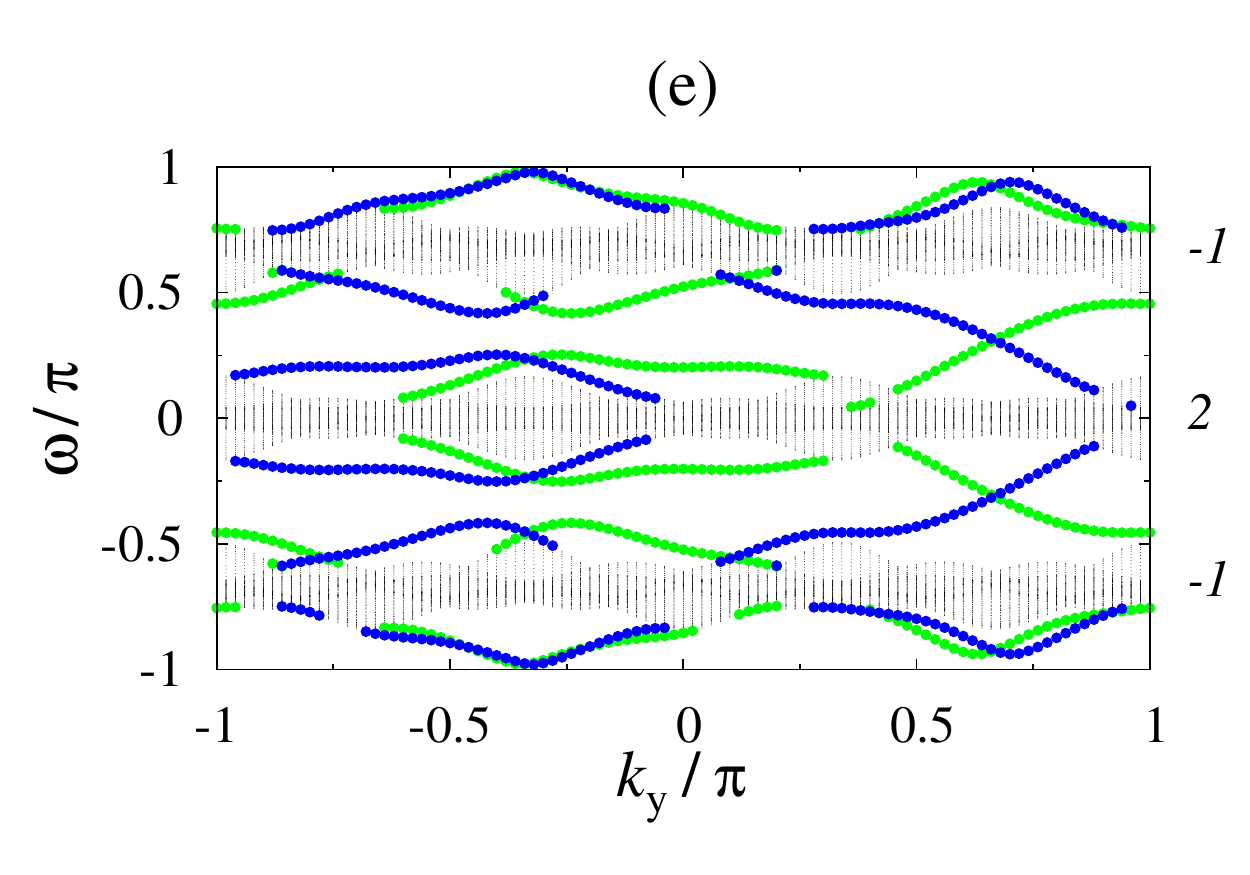} &
\includegraphics[height=!,width=7.5cm] {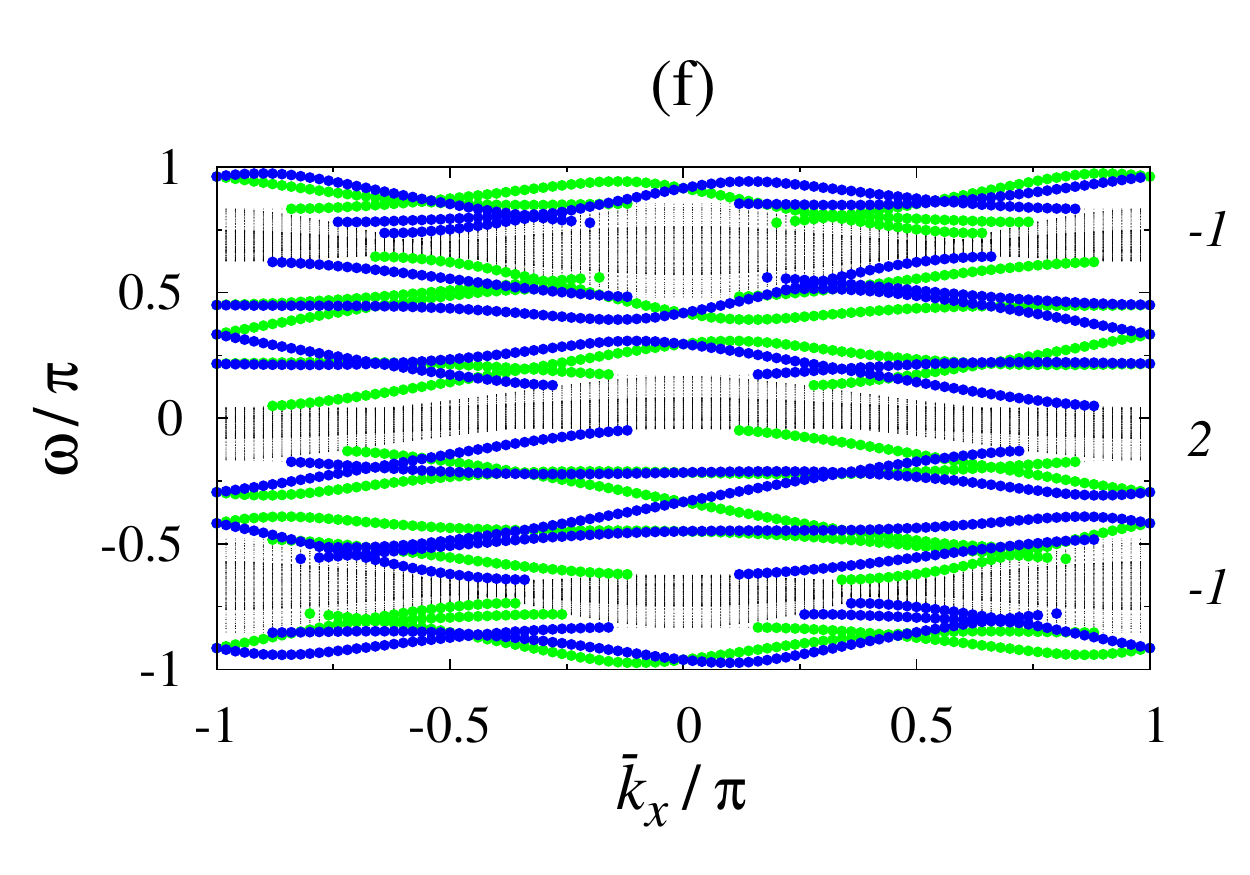}
\end{array}$
\end{center}
\caption{(color online). The QE spectra for the double kicked lattice model at $V=\pi/3$, $J_{2}=2\pi /3$ under open boundary conditions along $x$ ($y$) for $J_{1}=\pi, 2\pi, 3\pi$ are displayed in panels (a), (c), (e) [(b), (d), (f)] respectively. The Chern number bulk-boundary correspondence rule is obeyed regardless of choice of boundary. Panels (b), (d) and (f) are identical to those of the previous figure because the two models under open BCs along $y$ are still related by a unitary transformation. The edge states are plotted in the same fashion as in Fig. 1. } \label{ordkr-lattice-3band}
\end{figure}

We have also numerically computed similar quasienergy spectra under open BCs for the DKL model in Fig. \ref{ordkr-lattice-3band}. Here, we have chosen parameters so that the 1-1 correspondence \cite{KHMvsORDKR} between the DKL's eigenstates and the eigenstates of the KHL for the parameter choices in Fig. \ref{khm-lattice-3band} applies. We see again that the Chern number rule holds true regardless of boundary. The ACP modes fail to appear under open BCs along $x$ (see Figs. \ref{ordkr-lattice-3band}(a), (c), (e)) but can appear in some instance under open BCs along $y$ (see Fig.\ref{ordkr-lattice-3band}(d)). We note that for the parameter choices in Figs.\ref{ordkr-lattice-3band}(b), (d), (f) and \ref{khm-lattice-3band}(b), (d), (f) respectively, the DKL and KHL models are related by parameter mapping plus a unitary transformation, thus causing their overall spectra to be identical. This should \emph{not} be taken to mean that the DKL and KHL are actually the same system, for they are clearly quite different physically. We have calculated the quasienergy spectra of both models over a range of the $(R,J,J_{1},J_{2})$ parameters and found that the ACP modes, once present, do not disappear unless a band-touching occurs. This numerically suggests that these modes are of a topological nature.

\subsection{Symmetry Analysis}

Having presented our numerical data on the two models, we proceed to analytically study the symmetries of the Floquet operators. We build on the work of Ref. \cite{ZhaoKickedHall} and discuss the symmetry conditions related with the ACP modes in both systems. When open BCs are taken along $x$, the Floquet operators may be written as
\begin{equation}
U' = \sum_{k_{y}} [U' (k_{y})] \otimes \ket{k_{y}}\bra{k_{y}},
\end{equation}
where $U'$ refers generically to Floquet operators of either system and when necessary to distinguish which one we are referring to, we shall include appropriate subscripts. Here, $[U' (k_{y})]$ is a square matrix of dimension $L_{x} \times L_{x}$. We define similar notation when open BCs are taken along $y$, in which case the Floquet operators are written again generically as
\begin{equation}
U' = \sum_{\bar{k}_{x}} [U' (\bar{k}_{x})] \otimes \ket{\bar{k}_{x}}\bra{\bar{k}_{x}},
\end{equation}
with $[U' (\bar{k}_{x})]$ now being a matrix of dimension $3L_{y} \times 3L_{y}$.

With the above notation in place, we proceed with our analysis. It may be shown that under open BCs along $x$ and periodic BCs along $y$, the KHL model obeys
\begin{equation}
\Gamma_{\text{KH}} \sum_{k_{y}} [U'_{\text{KHL}} (k_{y})] \otimes \ket{k_{y}}\bra{k_{y}} \Gamma_{\text{KH}} = \sum_{k_{y}} [U'_{\text{KHL}} (k_{y}-\pi)]^{-1} \otimes \ket{k_{y}-\pi}\bra{k_{y}-\pi}. \label{KHL-CS-OBC-ky}
\end{equation}
whereas the DKL model obeys
\begin{equation}
\Gamma_{\text{DK}} \sum_{k_{y}} [U'_{\text{DKL}} (k_{y})] \otimes \ket{k_{y}}\bra{k_{y}} \Gamma_{\text{DK}} = \sum_{k_{y}} [U'_{\text{DKL}} (k_{y})]^{-1} \otimes \ket{k_{y}}\bra{k_{y}}.  \label{DKL-CS-OBC-ky}
\end{equation}
Taking open BCs along $y$ and periodic BCs along $x$, the KHL model obeys
\begin{equation}
\Gamma_{\text{KH}} \sum_{\bar{k}_{x}} [U'_{\text{KHL}} (\bar{k}_{x})] \otimes \ket{\bar{k}_{x}}\bra{\bar{k}_{x}} \Gamma_{\text{KH}} =  \sum_{\bar{k}_{x}} [U'_{\text{KHL}} (\bar{k}_{x}-\pi)]^{-1} \otimes \ket{\bar{k}_{x}-\pi}\bra{\bar{k}_{x}-\pi}, \label{KHL-CS-OBC-kx}
\end{equation}
and the DKL model obeys
\begin{equation}
\Gamma_{\text{DK}}\sum_{\bar{k}_{x}} [U'_{\text{DKL}} (\bar{k}_{x})] \otimes \ket{\bar{k}_{x}}\bra{\bar{k}_{x}} \Gamma_{\text{DK}} = \sum_{\bar{k}_{x}} [U'_{\text{DKL}} (\bar{k}_{x}-\pi)]^{-1} \otimes \ket{\bar{k}_{x}-\pi}\bra{\bar{k}_{x}-\pi}. \label{DKL-CS-OBC-kx}
\end{equation}
To summarize the above four relations, let $k$ be a generic crystal momentum variable and refer to either $\bar{k}_{x}$ or $k_{y}$ when appropriate. We then see two different types of CS here. Firstly, we have Eq. (\ref{DKL-CS-OBC-ky}) where the CS operator transforms each momentum space Floquet operator at $k$ into its own inverse. We call this a Type I CS. Secondly, we have Eqs. (\ref{KHL-CS-OBC-ky}), (\ref{KHL-CS-OBC-kx}) and (\ref{DKL-CS-OBC-kx}) where the CS operator transforms each momentum space Floquet operator at $k$ into the inverse of the momentum space Floquet operator at $k-\pi$. We call this a Type II CS.

A comparison of Eqs. (\ref{KHL-CS-OBC-ky}-\ref{DKL-CS-OBC-kx}) and Figs. \ref{khm-lattice-3band} and \ref{ordkr-lattice-3band} reveals that whenever ACP modes are present in either model, the CS is always of Type II. As noted in Ref.\cite{ZhaoKickedHall}, the Type II CS requires that for any eigenstate with arbitrary quasienergy $\omega$ at $k_{y}$ ($\bar{k}_{x}$), there must also exist an eigenstate with quasienergy $-\omega$ at $k_{y}-\pi$ ($\bar{k}_{x}-\pi$), which then implies that chiral modes crossing the gap at $\pm \pi$ must come in pairs. However, this is not yet enough to guarantee that the ACP modes, once present, are indeed topological (i.e. that they cannot be eliminated unless a band-touching occurs). This is because it could conceivably happen that as the system parameters are tuned, the ACP modes could develop a crossing which subsequently opens a gap, giving rise to an avoided crossing of edge modes \cite{if-gap-opens}. Our numerics in Figs. \ref{khm-lattice-3band}(d) and \ref{ordkr-lattice-3band}(d) show that this does not in fact happen. Namely, the quasienergies of the ACP modes cross without opening a gap. In our numerics, we have verified over a range of parameters that this crossing is always preserved so long as no band-touching occurs in the gap, thus confirming that the ACP modes are topological. We note that fact that the ACP modes never open a gap hints at the existence of some underlying symmetry.

Moving on, we consider the Type I CS obeyed by the DKL model under open BCs along $x$. The Type I CS implies a reflection symmetry of the spectrum about the $\omega =0$ axis, so that the chiral modes, if present in the $\pm \pi$ gap, must come in pairs with a crossing at $\omega = \pm \pi$. We have performed a large number of numerical simulations which show that this does not result in ACP modes because the `would-be ACP modes' always repel one another at $\omega = \pm \pi$, as seen for instance in Figs. \ref{ordkr-lattice-3band}(c) and (e) where they turn back and rejoin the bulk rather than crossing the gap. It appears that, unlike the situation in Figs. \ref{khm-lattice-3band}(d) and \ref{ordkr-lattice-3band}(d), crossings at $\pm \pi$ are impossible and generically lead to avoided crossings. Hence, our numerics suggest that whether the system possesses Type I or Type II CS plays an important role in determining the presence of the ACP modes. We remark that these symmetry-based considerations here are only relevant to those ACP modes in the $\pm \pi$ gap and do not apply to the modes in the middle gaps of Fig. \ref{khm-lattice-3band}(e).

We also note that, as mentioned at the end of Sec. \ref{GASymmetries}B, the 3-band KHL and DKL models are related by a rearrangement of their eigenstates on the $(\bar{k}_{x},k_{y})$ BZ. The above results show that the ACP modes are destroyed by such rearrangement, whereas the usual chiral modes described by the Chern numbers are not. This appears to be another clue which might be useful for better understanding the conditions for the existence of ACP modes in future.

Summarizing our main contributions in this section, we have shown numerically that ACP modes first observed in the KHL model in Ref. \cite{ZhaoKickedHall} may also be found in several other situations. We have shown that these modes depend on the choice of boundary, suggesting that they are a weak topological effect \cite{Hatsugai-2013JpnSoc}. We have also discussed how the ACP modes appear to be related to a particular form of (Type II) CS operator. Our results suggest that the ACP modes may be ubiquitous in Floquet operators obeying Type II CS. Lastly, we have also shown that the ACP modes are not robust against a rearrangement of eigenstates on the crystal momentum BZ, an insight which may be useful for an improved understanding of these modes in future.

\section{Topological States in 2-band Cases}

We move on now to the 2-band spectra corresponding to cases where $V=\pi/2$ in Eq. (\ref{lattice-UORDKR-2D-Symm-frame}) and $b=\pi$ in Eq. (\ref{lattice-UKHM-2D-symm-frame}) respectively. In the first subsection, we study analytically the models' CS operators in the bulk by assuming periodic BCs along both $x$ and $y$. This bulk analysis predicts that the 2-band DKL should possess topological $0$ and $\pi$ edge modes, but only at open boundaries taken along the $x$ direction, whereas the 2-band KHL should not possess any topological edge modes along any open boundaries. This sensitivity to choice of edge orientation in the DKL model is characteristic of a WTI phase. In the second subsection, we evaluate the $1D$ invariants associated with the $0$ and $\pi$ modes in the DKL model in various parameter regimes. The main interesting observation to arise out of this is the existence of a large number of edge modes under certain parameter choices.

\subsection{Analysis of Chiral Symmetry in 2-band models}

Making use of Eq. (\ref{Discrete-FT}), the Floquet operators can be written in momentum space, taking the form
\begin{equation}
U=\sum _{\bar{k}_{x}, k_{y}} e^{-i H_{\mathrm{eff}} (\bar{k}_{x},k_{y})} \otimes \ket{\bar{k}_{x}} \bra{\bar{k}_{x}} \otimes \ket{k_{y}}\bra{k_{y}}, \label{U-sum-of-k}
\end{equation}
where
\begin{equation}
H_{\mathrm{eff}}(\bar{k}_{x},k_{y}) = {\bf h} (\bar{k}_{x},k_{y}) \cdot {\bf \sigma} \label{h-vector}
\end{equation}
is a $2 \times 2$ effective Hamiltonian describing transitions within the reciprocal sublattice index degree of freedom. Within each particular $(\bar{k}_{x},k_{y})$ subspace, $(H')_{1,1}$ ($(H')_{2,2}$) describes transitions from the $A$ ($B$) sublattice back onto itself, while $(H')_{1,2}$ ($(H')_{2,1}$) describes transitions from the $B$ ($A$) sublattice onto the $A$ ($B$) sublattice.

Next, using Eq. (\ref{Discrete-FT}) again, we write the two bulk CS operators in the same momentum representation, beginning with the DKL model. The DKL model's CS operator given in Eq. (\ref{CS-OP-DKL}) reads as
\begin{eqnarray}
\Gamma_{\text{DK}} &=& \sum_{\bar{k}_{x},k_{y}} e^{-i\frac{\pi}{2}}e^{i\frac{\pi}{2}\sigma_{z}} \otimes  \ket{\bar{k}_{x}}\bra{\bar{k}_{x}} \otimes \ket{k_{y}}\bra{k_{y}}. \label{CS-z-rotatn} \nonumber \\
\end{eqnarray}
Within each $(\bar{k}_{x},k_{y})$ subspace, we observe that the CS operator has the form of a rotation of Pauli vectors by an angle of $\pi$ about the $z$-axis (with an unimportant phase factor attached). Considering the DKL Floquet operator $U'_{\text{DKL}}$ written in the form of Eq. (\ref{U-sum-of-k}), the former observation then necessarily implies that the $z$-component of the ${\bf h}' (\bar{k}_{x},k_{y})$ vector must be zero for all values of $(\bar{k}_{x},k_{y})$. This implication is confirmed when we write out $U'_{\text{DKL}}$ in momentum space, which reads as
\begin{equation}
U'_{\text{DKL}}=\sum _{\bar{k}_{x}, k_{y}} e^{-i H'_{\mathrm{eff}} (\bar{k}_{x},k_{y})} \otimes \ket{\bar{k}_{x}} \bra{\bar{k}_{x}} \otimes \ket{k_{y}}\bra{k_{y}}, \label{DKL-sum-of-k}
\end{equation}
where
\begin{equation}
H'_{\mathrm{eff}}(\bar{k}_{x},k_{y}) = {\bf h}' (\bar{k}_{x},k_{y}) \cdot {\bf \sigma},
\end{equation}
and the explicit form of ${\bf h}' (\bar{k}_{x},k_{y})$ is given by
\begin{eqnarray}
{\bf h}' (\bar{k}_{x},k_{y}) &=& E(\bar{k}_{x},k_{y}) {\bf n}' (\bar{k}_{x},k_{y}), \nonumber \\
E(\bar{k}_{x},k_{y}) &=& \cos ^{-1} \left[ \cos (P) \cos (Q)\right], \nonumber \\
{\bf n}' (\bar{k}_{x},k_{y}) &=& \left[ n_{x}' (\bar{k}_{x},k_{y}), n_{y}' (\bar{k}_{x},k_{y}), 0 \right], \nonumber \\
n_{x}' (\bar{k}_{x},k_{y}) &=& \frac{\cos(\frac{\bar{k}_{x}}{2}) \sin(P) \cos(Q) -\sin(\frac{\bar{k}_{x}}{2})\sin(Q)}{\sin(E(\bar{k}_{x},k_{y}))}, \nonumber \\
n_{y}' (\bar{k}_{x},k_{y}) &=& \frac{-\sin(\frac{\bar{k}_{x}}{2})\sin(P)\cos(Q) - \cos(\frac{\bar{k}_{x}}{2})\sin(Q)}{\sin(E(\bar{k}_{x},k_{y}))}, \nonumber \\
P & \equiv & J_{2} \cos \left(\frac{\bar{k}_{x}}{2} \right), \nonumber \\
Q & \equiv & J_{1} \cos \left(k_{y} +\frac{\bar{k}_{x}}{2} \right). \label{H-prime}
\end{eqnarray}
For each fixed $\bar{k}_{x}$ ($k_{y}$) value, we have an \emph{effective} 1D Floquet system (but we stress that the DKL is physically a bona fide 2D system) whose ${\bf h}'$ vector we can track as $k_{y}$ ($\bar{k}_{x}$) is scanned across the BZ. Because this vector always lies in one plane for all values of $\bar{k}_{x}$ and $k_{y}$, it is possible to define topological winding numbers counting the number of circles ${\bf h}' (\bar{k}_{x},k_{y})$ traces around the origin as $\bar{k}_{x}$ ($k_{y}$) is tuned from $-\pi$ to $\pi$. A very similar situation occurs in graphene\cite{Hatsu-topo-graphene,Hatsu-CS-graphene}. It is clear that ${\bf h''} (\bar{k}_{x},k_{y})$ corresponding to the Floquet operator in the second symmetric time frame $U''_{\text{DKL}}$ also lies entirely in the $x$-$y$ plane since it shares the same CS operator (see Appendix B for the explicit form of $U''_{\text{DKL}}$ in momentum space) and a similar winding number may be defined as well.

These winding numbers are related with the number of topologically protected $0$ and $\pi$ quasienergy edge modes in 1D lattices \cite{AsbothBBC,AsbothBBC1DSys}. Here, in our 2D DKL model, the winding number as $\bar{k}_{x}$ ($k_{y}$) is tuned from $-\pi$ to $\pi$ is associated with the number of $0$ and $\pi$ modes which occur along an open boundary along $x$ ($y$) within each $k_{y}$ ($\bar{k}_{x}$) subspace. The occurrence of non-zero winding numbers generally means that edge modes will be present (see Sec. V.B for further details and the actual values of these winding numbers). We note that the DKL model is described not by one winding number but by an entire ensemble of them. Each $\bar{k}_{x}$ ($k_{y}$) at which no band-touching takes place hosts a single winding number. As we shall see later, the 2-band DKL model possesses non-zero winding numbers and thus displays weak Floquet topological insulating phases where $0$ and $\pi$ quasienergy modes are found only along some edges but not others.

\begin{figure}
\begin{center}$
\begin{array}{cc}
\includegraphics[height=!,width=7.5cm] {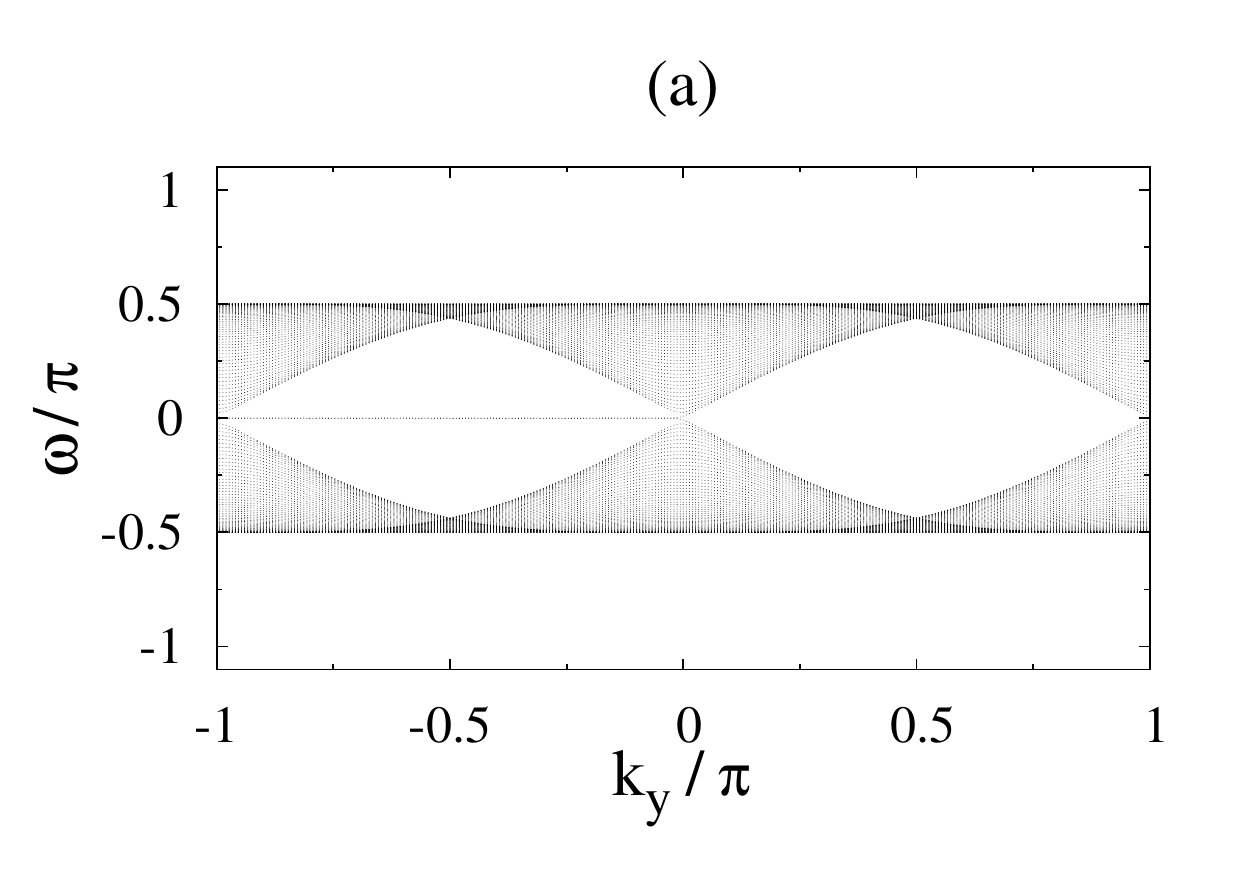}  &
\includegraphics[height=!,width=7.5cm] {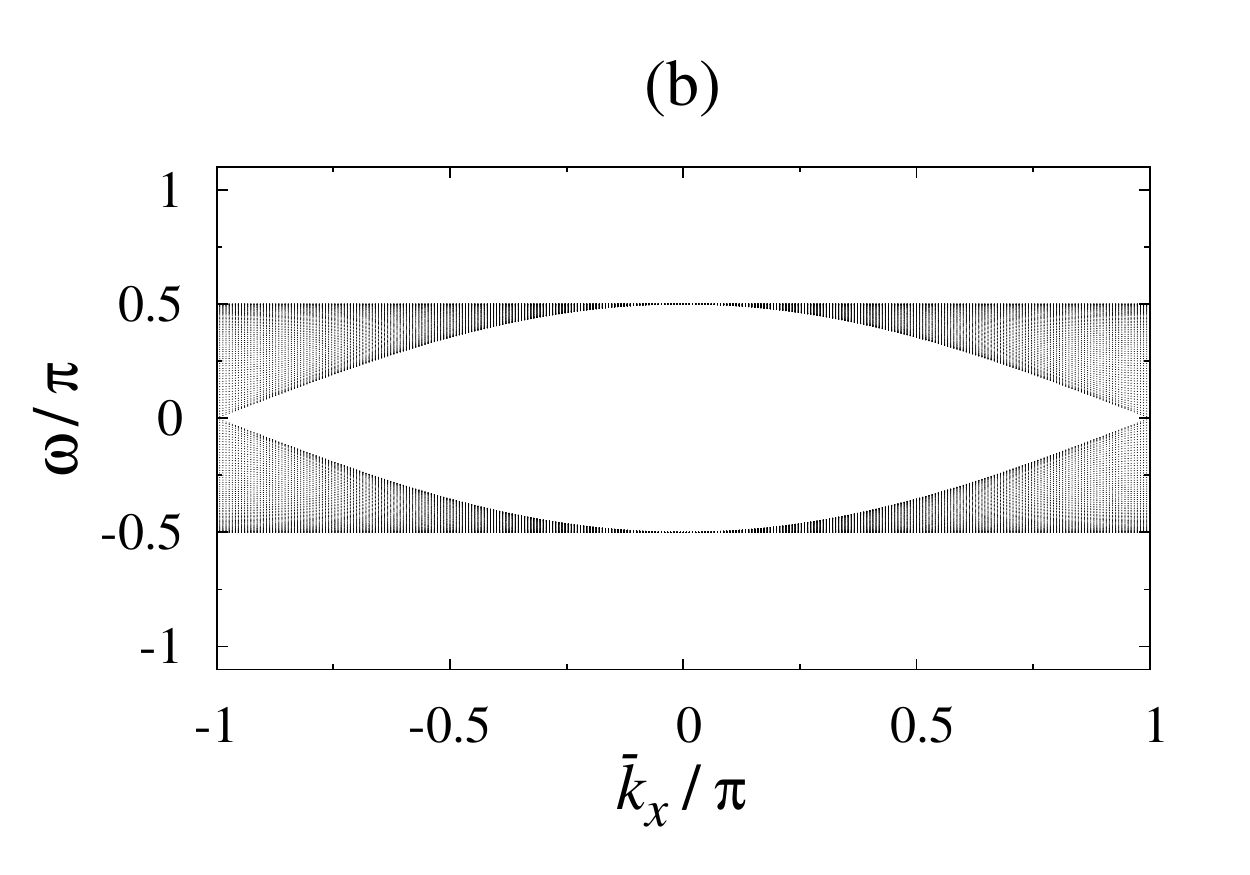} \\
\includegraphics[height=!,width=7.5cm] {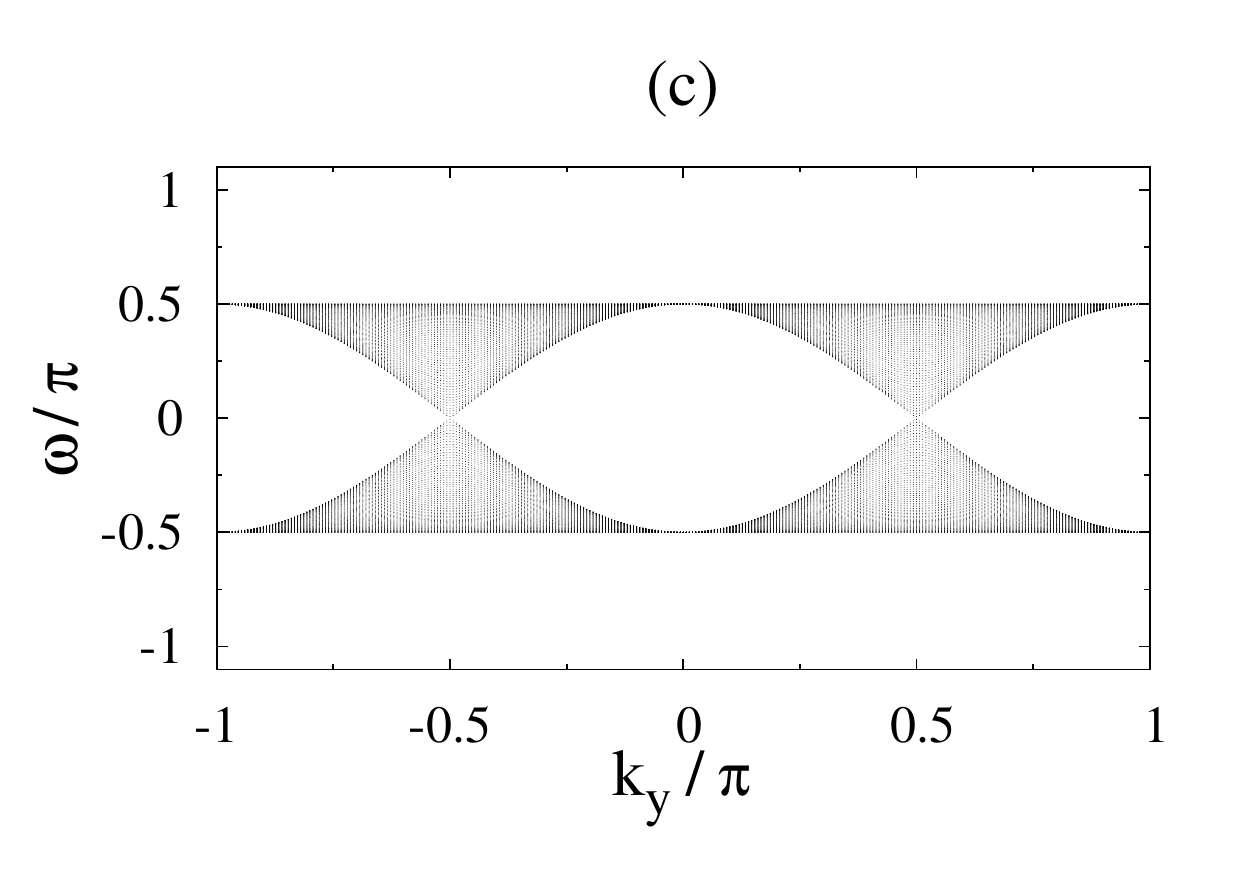}  &
\includegraphics[height=!,width=7.5cm] {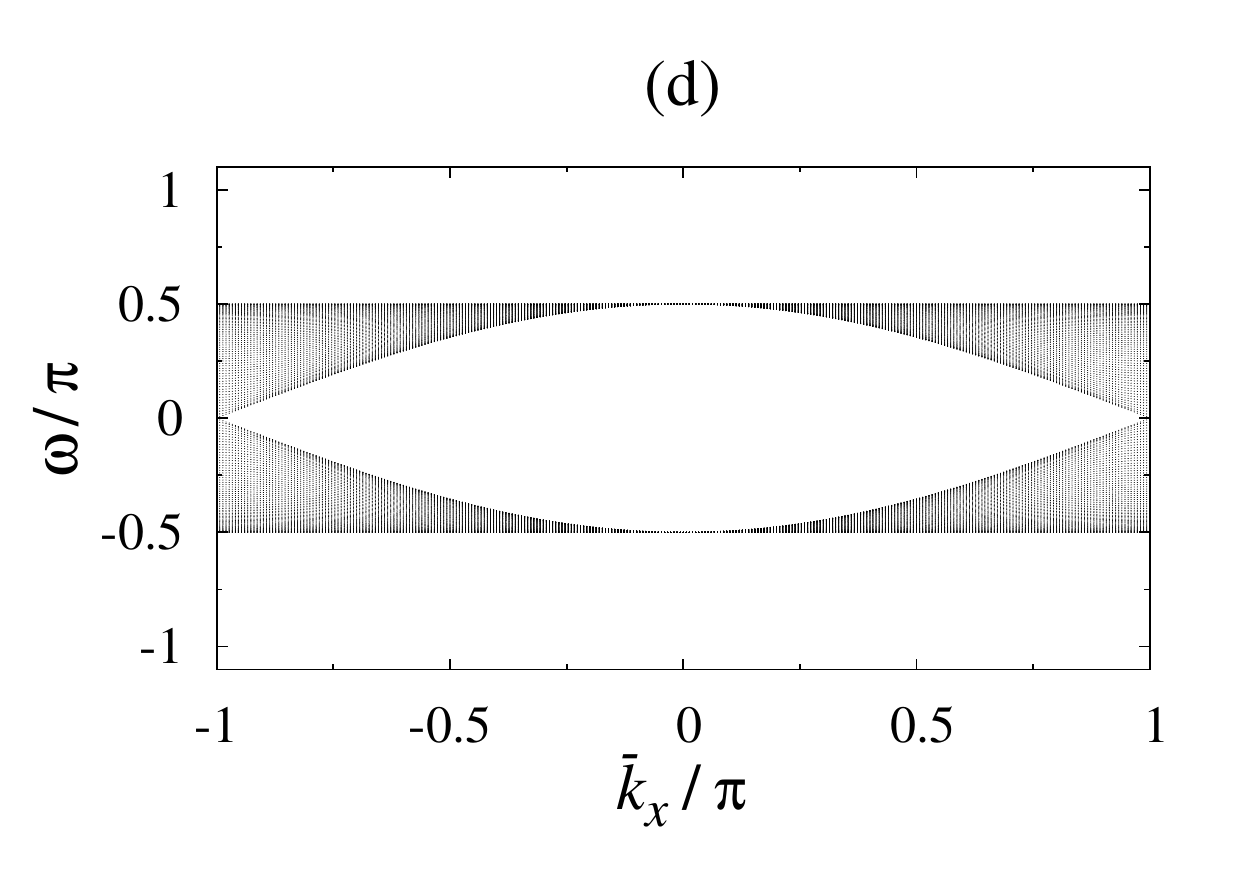} \\
\end{array}$
\end{center}
\caption{The QE spectra for the DKL (KHL) model for $V=\pi/2$, $J_{1}=J_{2}=0.5 \pi$ [$b=\pi$, $R=J=0.5\pi$] under open BCs along $x$ and $y$ are shown in panels (a) and (b) [(c) and (d)] respectively. No topological edge modes appear in the case of the KHL, as expected by its lack of topological winding numbers. Topologically protected $0$ modes appear in the case of the DKL but only for open BCs along $x$. The reason for this is explained later in the text.} \label{ordkr-vs-khm-2bands}
\end{figure}

Next, we analyse the KHL in the bulk. The KHL's CS operator in Eq. (\ref{CS-OP-KHL}) in $(\bar{k}_{x},k_{y})$ representation reads as
\begin{equation}
\Gamma_{\text{KHL}} = \sum_{\bar{k}_{x},k_{y}} e^{-i\frac{\pi}{2}}e^{i\frac{\pi}{2}\sigma_{z}} \otimes  \ket{\bar{k}_{x}}\bra{\bar{k}_{x}} \otimes \ket{k_{y}-\pi}\bra{k_{y}}.
\end{equation}
The CS operator clearly does not conserve $k_{y}$. It is thus impossible to show using this CS operator that we can fix one crystal momentum and obtain an effective 1D Floquet operator with corresponding ${\bf h}$ vector that lies purely in one plane. There are thus no well-defined topological winding numbers in the 2-band KHL.
This non-existence of well-defined winding numbers in the 2-band KHL is consistent with our numerical observation (see below) that it does not host any topological edge modes.
The DKL model, on the other hand, may or may not possess such edge modes, depending on the values of its winding numbers. We will evaluate these numbers in the next subsection, but for now we show in Fig. \ref{ordkr-vs-khm-2bands} a typical example of the spectra in both models under open BCs along $x$ and $y$.  We see that, as expected, the DKL possesses $0$ quasienergy edge modes while the KHL does not. We defer the introduction and calculation of the winding numbers associated with the DKL edge modes to the following subsection, because these winding numbers are defined with respect to the bulk (i.e., under periodic BCs). For now, we wish to discuss how under open BCs, the CS operator of the DKL also provides a mechanism for topological protection of the edge modes, whereas that of the KHL does not. This coexistence of protected edge modes defined under open BCs with topologically invariant winding numbers defined under periodic BCs is a typical example of the bulk-boundary correspondence principle.

We consider the spectra of both models under open BCs along $x$. In the case of the DKL, the CS operator $\Gamma_{\text{DK}}$ transforms $U'_{\text{DKL}}(k_{y})$ into $U^{'\dagger}_{\text{DKL}}(k_{y})$. On the other hand, in the case of the KHL, the different CS operator $\Gamma_{\text{KH}}$ transforms $U'_{\text{KHL}}(k_{y})$ into $U^{'\dagger}_{\text{KHL}}(k_{y}-\pi)$.
Because of this, so long as CS is maintained in the DKL model, if there exists an eigenstate with quasienergy $\omega$ at some $k_{y}$, there must also exist an eigenstate with quasienergy $-\omega$ at \textit{the same} $k_{y}$. For $\omega=0$ (or $\pi$) quasienergies, these quasienergies could correspond to one and the same eigenstate (note that $\pi$ and $-\pi$ are the same in the quasienergy BZ). This allows us to explain the topological protection in the DKL model in the following intuitive but non-rigorous manner. If at some $k_{y}$ there is a single eigenstate with quasienergy $\omega=0$ ($\pi$) within a gap, this state is not allowed to move away from $0$ ($\pi$) quasienergy under any CS-preserving perturbation \cite{KitagawaNatComm} for the simple reason that a single state cannot suddenly split into two under continuous change of parameters. This constitutes a topological protection of \textit{single} $0$ and $\pi$ quasienergy edge states in the DKL model. This argument is however unable to explain whether or not multiple $0$ (or $\pi$) modes may be simultaneously protected, as we can always imagine say a pair of $0$ modes simultaneously moving away from $0$ in opposite directions, thus preserving the chiral symmetry of the spectrum. It turns out that multiple modes may indeed be simultaneously protected in the DKL model. A proof of this is provided in Appendix A based on a very similar analysis in Ref. \cite{KitagawaNatComm}.
The situation is rather different for the $U'_{\text{KHL}}$ spectrum. In this case, if there exists an eigenstate with quasienergy $\omega$ at some $k_{y}$, then the CS condition only requires that there must also exist an eigenstate with quasienergy $-\omega$ at $k_{y}-\pi$. Because the chiral symmetry partner lies at a \textit{different} value of $k_{y}$, it is thus guaranteed to be a distinct eigenstate, so our previous ``thought scenario" arguing how a single state cannot split into two no longer applies. Hence, the presence of CS here does not offer a mechanism towards a topological protection of states with quasienergies $0$ or $\pi$. We note that the difference in edge state behaviour between the two models ultimately stems from the fact that $\Gamma_{\text{KH}}$ has less translational symmetry than the KHL Floquet operator, whereas $\Gamma_{\text{DK}}$ has the same translational symmetry as the DKL Floquet operator (i.e., periodic along $y$ over every $1$ lattice site).

Taking into account the above statements and the results of the previous section (cf. Eqs. (\ref{KHL-CS-OBC-ky})-(\ref{DKL-CS-OBC-kx})), we make the following observation. The existence of a CS operator with less translational symmetry than its Floquet operator gives rise to the ACP modes in 3-band cases, whereas in 2-band cases, this causes the Floquet operator to be topologically trivial. Conversely, the existence of a CS operator with the same translational symmetry as its Floquet operator does not allow for the existence of ACP modes in 3-band cases, but it allows for the existence of topologically protected $0$ and $\pi$ quasienergy edge modes in 2-band cases.

\subsection{Topological invariants for the bulk spectrum of DKL}

We discussed the topological protection of the edge modes of the 2-band DKL in the context of open BCs in the previous subsection. Here, we show that under periodic BCs, this corresponds to the existence of non-zero winding numbers, as expected based on the bulk-boundary correspondence principle. {We do not characterize the 2-band models in terms of Chern numbers like we did for the 3-band cases due to the following two reasons. Firstly, the 2-band Floquet spectra always have band-touching points, making it impossible to define each band's individual Chern number. Secondly, though adding new terms to the Floquet operators to open up a gap should be possible, it is unclear whether there exist such gap-opening terms which will not break the intriguing eigenstate mapping between the two models.}

Since we will make use of results from the theory in Refs. \cite{AsbothBBC,AsbothBBC1DSys}, we now recap them briefly. Assume we are given a 1D driven system described by a 2-band effective Hamiltonian (cf Eq. (\ref{U-sum-of-k})) corresponding to its Floquet operator which possesses CS. Corresponding to the two symmetric time frames, one then obtains vectors $\bf{h}'(k)$ and $\bf{h}''(k)$ in the same sense as Eq. (\ref{h-vector}), albeit in 1D, where $k$ here refers to a generic 1D crystal momentum. The vectors $\bf{h}'(k)$ and $\bf{h}''(k)$ possess winding numbers $\nu '$ and $\nu ''$ respectively, which count the number of times each vector encircles the origin as $k$ is scanned across one period of the Brillouin zone. Refs. \cite{AsbothBBC,AsbothBBC1DSys} showed that under open BCs, there exist at each boundary precisely $\nu_{0}$ ($\nu_{\pi}$) topologically protected $0$ ($\pi$) quasienergy modes, which are related to the aforementioned bulk winding numbers via \cite{AsbothBBC1DSys}
\begin{equation}
\nu_{0} = (\nu' + \nu'')/2,
\end{equation}
and
\begin{equation}
\nu_{\pi} = (\nu' - \nu'')/2.
\end{equation}

\begin{figure}
\begin{center}
\includegraphics[trim=4cm 4cm 4cm 4cm,clip=true,height=7cm,width=!] {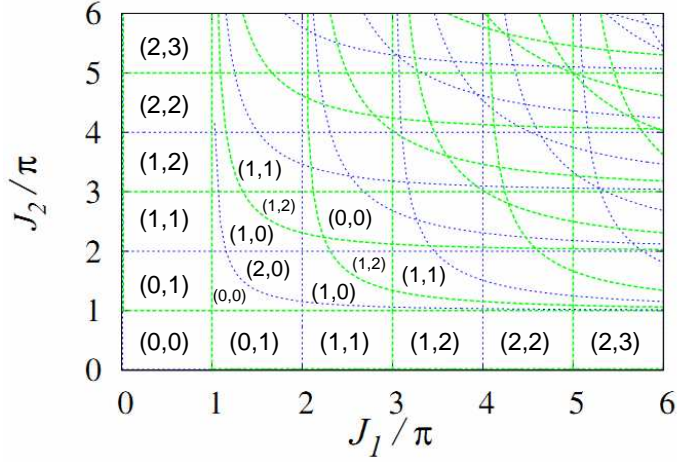}
\end{center}
\caption{(color online). Phase transition lines of the DKL model in the $(J_{1} , J_{2})$ space for $k_{y}=\pi/2$. Gap closures at $\omega=0 (\pi)$ are marked with a blue-dashed (green-dashed). The $(\nu_{0}, \nu_{\pi})$ numbers signifying the number of $0$ and $\pi$ modes respectively at each edge (under open boundary conditions) are indicated within each region of the parameter space. Note that the number of edge modes seems to increase without bound when we fix $J_{1}$  ($J_{2}$) at $0.5\pi$ and increase $J_{2}$ ($J_{1}$).} \label{phase-diagram}
\end{figure}

We now apply the above to our 2D 2-band DKL model. We have found through extensive numerical simulations that for open BCs along $y$, the DKL model does not host any edge modes. Consistent with this finding, the associated winding numbers $\nu' (\bar{k}_{x})$ and $\nu''(\bar{k}_{x})$ for all $\bar{k}_{x}$ are always zero, so that $\nu_{0}$ and $\nu_{\pi}$ are necessarily always zero as well. There are thus no topological edge states under open BCs along $y$. The same is not true for open BCs along $x$ as the winding numbers $\nu' (\bar{k}_{y})$ and $\nu''(\bar{k}_{y})$ take non-zero values and our numerics indicate the existence of $0$ and $\pi$ quasienergy modes. Hence, the 2-band DKL is indeed a weak Floquet topological insulator. We present in Fig. \ref{phase-diagram} below the values of $(\nu_{0}, \nu_{\pi})$ at $k_{y}=\pi/2$ over a large range of $(J_{1} , J_{2})$ values of the DKL model.
The topological phase diagrams for all other $k_{y} \neq 0,\pi$ can be easily obtained and are similar to Fig. \ref{phase-diagram}, differing only by some shifts of the transition lines.  The phase diagram is seen to possess a wide variety of different topological phases. In particular, for fixed $J_{1}= \pi/2$ (but any $0 < J_{1} < \pi$ will also suffice), as we increase the value of $J_{2}$, we will pass through alternate gap closures at $\omega=0$ and $\omega=\pi$. With each of these closures, $\nu''$ increases by $1$ while $\nu'$ alternates between $-1$ and 0. This pattern seems to carry on ad infinitum, meaning that the number of $\omega=0$ and $\omega=\pi$ topologically protected edge modes at each boundary with the vacuum, given by $\nu_{0} = (\nu'+\nu'')/2$ and $\nu_{\pi} = (\nu'-\nu'')/2$ respectively, will become very large as $J_{2}$ becomes large. A similar situation happens if we fix $J_{2} = \pi /2 $ and increase $J_{1}$.

An especially interesting feature is that, as we can see from Fig. \ref{phase-diagram}, the phase transition lines do not occupy the parameter space densely along the $J_{1}= 0.5 \pi $ ($J_{2}= 0.5 \pi $) line no matter how large $J_{2}$ ($J_{1}$) becomes, unlike in the regions in the upper right corner of Fig. \ref{phase-diagram} where the phase transition lines become increasingly dense as both $J_{1}$ and $J_{2}$ increase to large values. This means that even if the actual $J_{1}$ and $J_{2}$ values in an experiment are shifted due to reasonably small imperfections, the system does not undergo a phase transition and the $0$ and $\pi$ modes will thus persist. The model may thus be very well-sutied for realizing a large number of topologically protected edge modes, which might be useful for quantum information applications\cite{BarangerPRL,ManyMajoranasPRB}.

\begin{figure}
\begin{center}$
\begin{array}{cc}
\includegraphics[height=!,width=7.5cm] {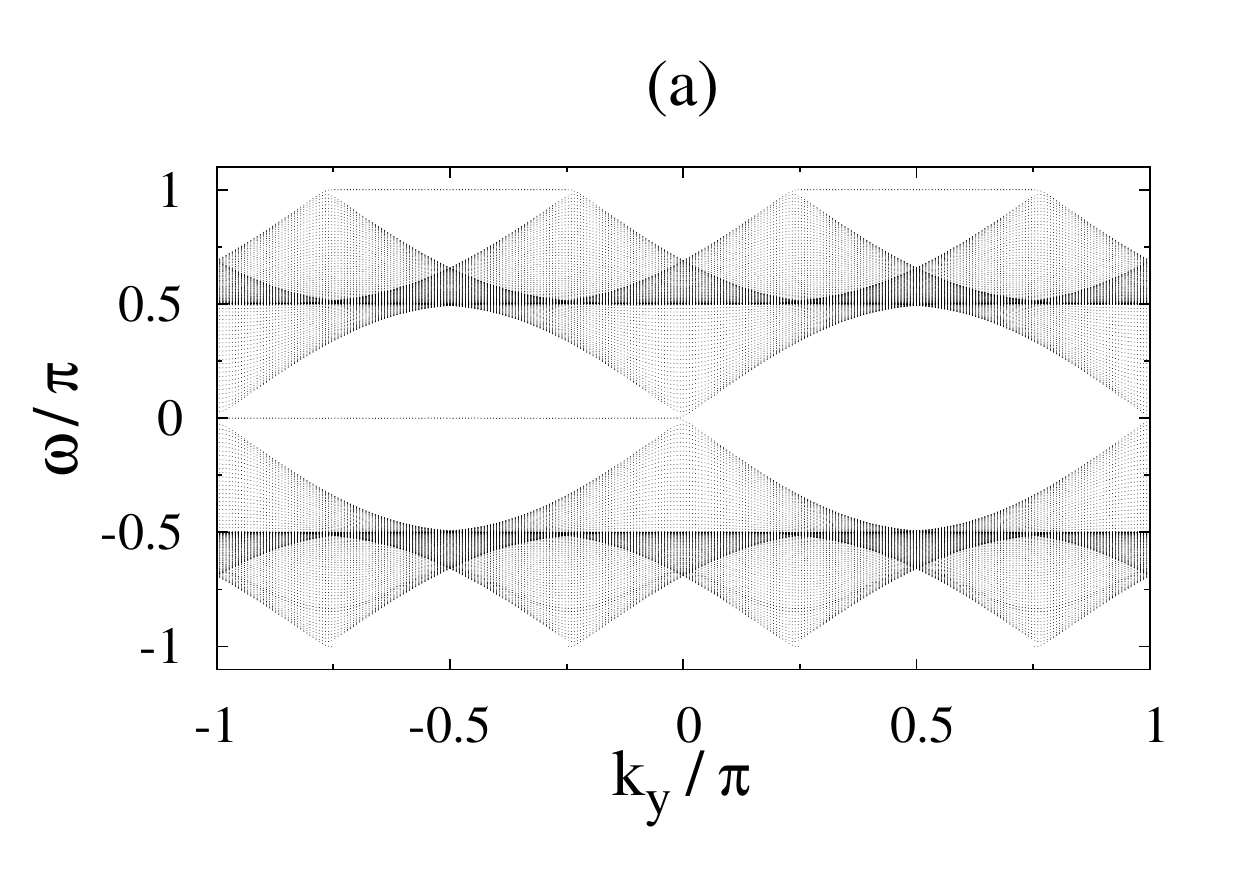}  &
\includegraphics[height=!,width=7.5cm] {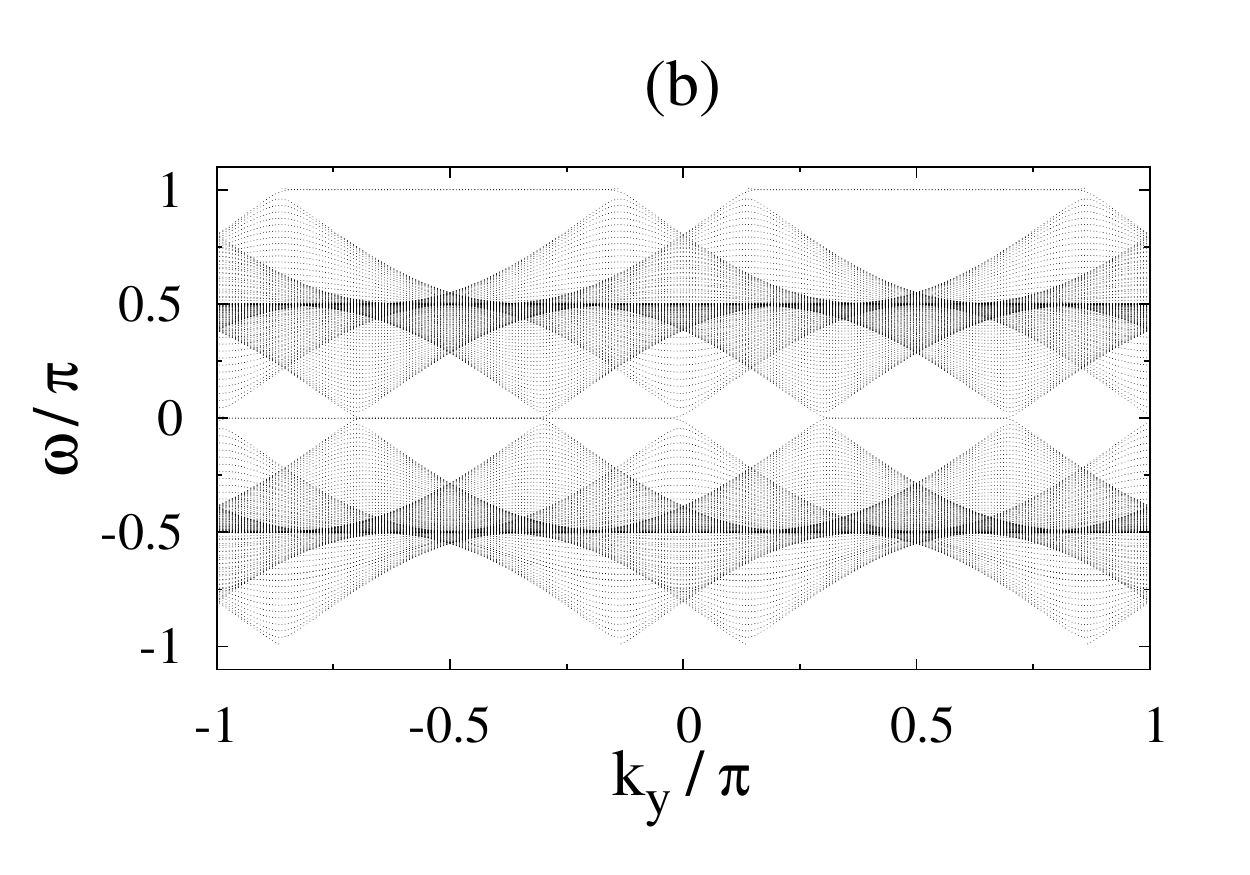} \\
\includegraphics[height=!,width=7.5cm] {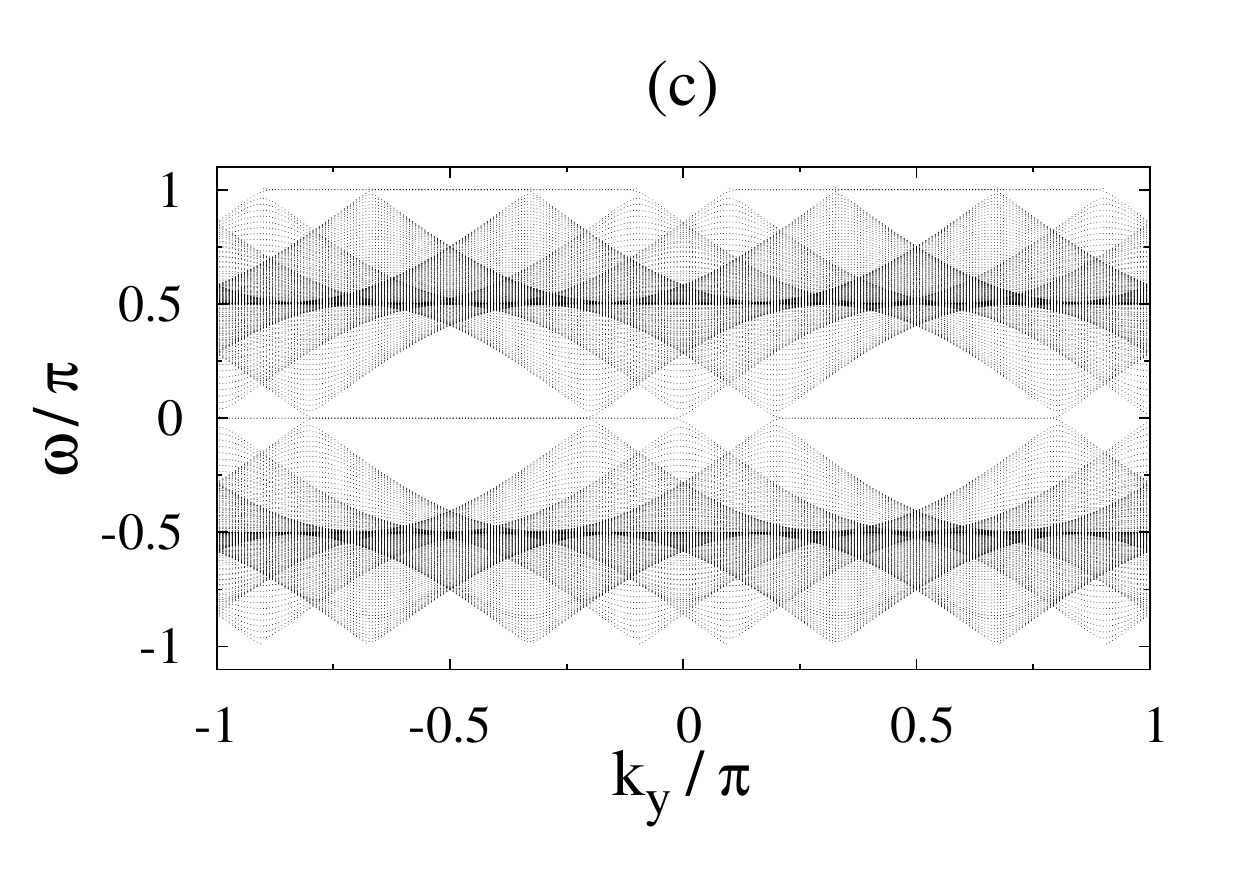} &
\includegraphics[height=!,width=7.5cm] {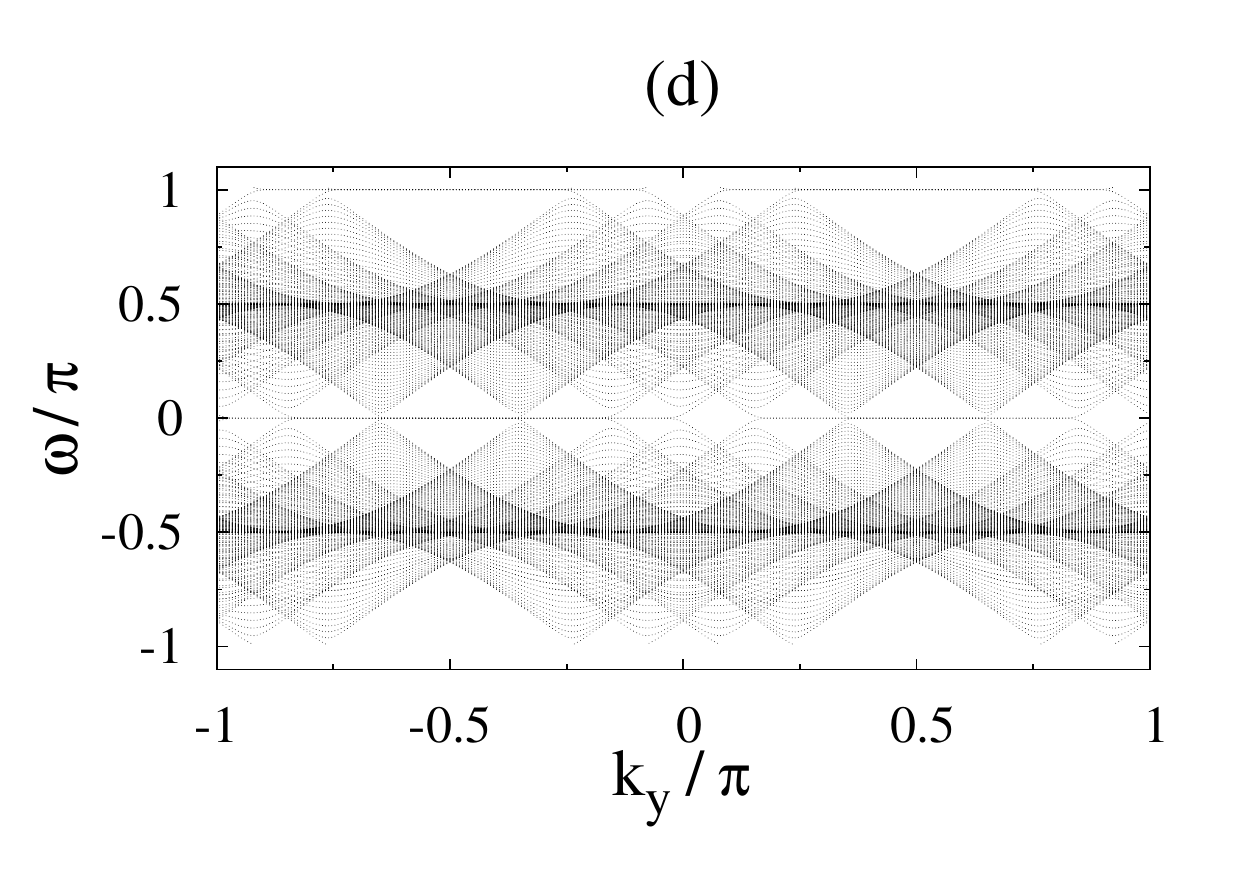}
\end{array}$
\end{center}
\caption{The QE spectra for the DKL model as a function of $k_{y}$ at $V=\pi/2$, $J_{1}=0.5 \pi$, and (a) $J_{2}=1.5\pi$, (b) $J_{2}=2.5\pi $, (c) $J_{2} = 3.5\pi$, (d) $J_{2} = 4.5\pi$. We see a proliferation of Dirac-like points at $\omega=0$ and $\omega=\pm \pi$ as $J_{2}$ increases.} \label{ordkr-lattice-2band}
\end{figure}

De-specializing away from the $k_{y} = 0.5 \pi$ case, we consider in Fig. \ref{ordkr-lattice-2band} the quasienergy spectrum under open BCs along $x$ over the whole $k_{y}$ BZ as $J_{2}$ increases with $J_{1}$ fixed at $0.5\pi$ (note also the $\pi$ quasienergy edge modes together with the 0 quasienergy edge modes). Firstly, we see that the topological $0$ and $\pi$ modes are present over an increasingly large interval of the BZ as $J_{2}$ increases. This shows that the winding numbers at all $k_{y}$ (and not just those at $0.5\pi$) generally increase as $J_{2}$ increases along the $J_{1}=0.5\pi$ line in parameter space. We also note that an increasing number of quasienergy Dirac cones as $J_{1}$ increases. Consider what happens when a phase transition line of the form $J_{2}= (2m+1) \pi, m\in \mathds{Z} $ is crossed. When $J_{2}= (2m+1) \pi$, a new cone forms at $k_{y} = \pm 0.5 \pi, \omega = \pi$. As $J_{2}$ increases further, the two cones do not vanish. Instead, each one splits into two and moves off to either side. Hence, we now have four more cones than we did before crossing the phase transition line. A similar sequence of events occurs when a $J_{2} =  2m \pi, m \in \mathds{Z}$ line is crossed. New Dirac cones occur at $k_{y} = \pm 0.5 \pi, \omega=0$ when $J_{2} = 2m\pi$ and split off into two upon further increase of $J_{2}$, again resulting in the presence of four more Dirac-like points than before the phase transition line was crossed. Hence, as $J_{2}$ is increased along the line $J_{1}=0.5\pi$, the number of Dirac cones increases rapidly. Since the DKL Floquet operator's quasienergy spectrum corresponds to the energy spectrum of an associated effective Hamiltonian $H_{\text{eff}}$ via Eq. (\ref{Heff-defn}), this proliferation of Dirac cones may be useful for simulating Hamiltonians with a tunable number of Dirac cones, a subject of considerable theoretical and experimental interest\cite{EsslingerNature2012,MontambauxPhysicaB,HasegawaPRB,SticletLongHoppingPRB2013,HouPRB2014,Tashima2014}. We note from Eq. (\ref{ORDKR-periods-2D}) that all one needs to do in order to effectively increase $J_{1}$ or $J_{2}$ in $U'_{\text{DKL}}$ is to increase the two time intervals during which $J(t) \neq 0$, an experimentally rather straightforward task.

In Refs. \cite{HasegawaPRB,SticletLongHoppingPRB2013,HouPRB2014}, the appearance of new Dirac cones was due to increasing either the hopping strength or hopping range in a static Hamiltonian. We point out that by increasing $J_{1}$ and $J_{2}$ here, we are effectively simulating a static Hamiltonian with long-range hopping\cite{ManyMajoranasPRB}. To see this, note the effective Hamiltonian is defined via
\begin{equation}
U_{\text{DKL}} \equiv e^{-i \hat{H}_{\text{eff}}}.
\end{equation}
The Floquet operator is given by the concatenation of four exponential operators as seen in Eq. (\ref{lattice-UORDKR-ky}). Each exponential operator does not commute with the exponential operator on either side of it. Hence, in order to obtain $\hat{H}_{\text{eff}}$, one must apply the Baker-Campbell-Hausdorff (BCH) formula to each pair of adjacent exponential operators repeatedly until we finally are left with only one exponential operator. Now, by making use of the BCH formula, we see that given three arbitrary operators $\hat{X},\hat{Y}$ and $\hat{Z}$ related via
\begin{equation}
e^{-i\hat{Z}} \equiv e^{-ic_{1}\hat{X}}e^{-ic_{2}\hat{Y}},
\end{equation}
where $c_{1},c_{2}$ are c-numbers, the operator $\hat{Z}$ is given by
\begin{equation}
\hat{Z} = c_{1} \hat{X} + c_{2} \hat{Y} - \frac{ic_{1}c_{2}}{2} [\hat{X},\hat{Y}] - \frac{c_{1}c_{2}}{12}[c_{1}\hat{X}-c_{2}\hat{Y},[\hat{X},\hat{Y}]] + \cdots \hspace{1mm}.
\end{equation}
Due to the infinite series of nested commutators of $\hat{X}$ and $\hat{Y}$, we see that $\hat{Z}$ may contain terms of longer-range hopping than those present in both $\hat{X}$ and $\hat{Y}$ individually. The larger the values of $c_{1}$ and $c_{2}$, the more nested commutator terms will play a significant role in $\hat{Z}$. Applying this in the context of the problem at hand, we conclude that $\hat{H}_{\text{eff}}$ will contain longer-range hopping terms beyond the nearest-neighbour hopping terms seen in Eq. (\ref{LDR2D}). Larger values of $J_{1}$ and $J_{2}$ will then lead to longer-range hopping in $\hat{H}_{\text{eff}}$. As we saw earlier, the Floquet operator of the DKL model possesses CS regardless of the values of $J_{1}$ and $J_{2}$. Hence, by increasing these values, we are able to simulate an effective chiral symmetric Hamiltonian with very long-range hopping. As mentioned earlier, increasing $J_{1}$ and $J_{2}$ is achieved simply by prolonging the `hopping' part of the period.

{Summarizing this section, we have found that, despite the existence of a  mapping  between the DKL and KHL models, they possess different topological behaviour in the two-band cases as well. Namely, the DKL possesses weak Floquet topological edge states but the KHL does not.
These results reinforce the observation from the three-band cases investigated in Sec.~IV.
 That is, a re-arrangement of eigenstates in the crystal momentum BZ can create or destroy weak topological effects.  We have also found that the 2-band DKL is able to host a large number of topological modes while at the same time generating a large number of Dirac cones in a manner which is experimentally appealing.}

\section{Conclusion}

We have studied two chiral symmetric driven 2D quantum systems and demonstrated theoretically that they host weak topological edge states. In the 3-band cases, we found in both models that anomalous counter-propagating (ACP) chiral edge modes exist only along certain boundaries and persist over a wide parameter range, thus suggesting that these are a weak topological effect. If this is the case, there ought to exist a weak topological invariant associated with their existence. At the time of writing, such an invariant has yet to be discovered. Our results suggest that a crucial ingredient for the topological protection of these ACP modes is the existence of a chiral symmetry (CS) operator which maps each Floquet operator in momentum space at crystal momentum $k$ onto its inverse at $k-\pi$ \cite{ZhaoKickedHall}, where $k$ here denotes a generic crystal momentum variable and the system is studied in a cylinder geometry, meaning that periodic BCs taken along one direction and open BCs along the other.

In the 2-band cases, the situation is somewhat reversed. The existence of a CS operator transforming each momentum space Floquet operator at $k$ into the inverse of that at $k-\pi$ does not protect the existence of edge modes. Instead, the existence of a CS operator mapping each Floquet operator at $k$ onto its inverse at the same $k$ is required to topologically protect edge modes. We have also found that an arbitrarily large number of topological $0$ and $\pi$ quasienergy edge modes may be generated by simply increasing the duration of the hopping stages within each time period of the 2-band DKL Hamiltonian.  These modes could be useful for future quantum information applications \cite{BarangerPRL}. Finally, we also showed that this gives rise to a proliferation of Dirac cones in the quasienergy spectrum, a finding which may be useful for simulating static chiral-symmetric Hamiltonians with many Dirac cones.

We have also emphasized that the Floquet eigenstates of the two dynamical models studied in this work have an interesting correspondence and hence differ (up to a unitary transformation) in their arrangement on the BZ. Both our 3-band and 2-band results (as prototypical representatives of the even-band and odd-band cases) indicate that weak topological effects depend not just on the nature of the set of eigenstates associated with a physical system, but also on the arrangement of these states on the BZ.

For possible experimental realizations of our findings here, we note that photonic setups are increasingly establishing themselves as a versatile setup for simulating topological quantum phases \cite{NatureFTI,PasekPRB}. Another possible avenue to consider would be optical lattice setups \cite{RoatiHarper,MReichl,GoldmanDirectImaging}. We note that the authors of Ref. \cite{ZhaoKickedHall} suggest that the 3-band KHL model may possibly be realized by making use of artificial magnetic field techniques \cite{ArtificialMagFields} or by introducing complex tunnelling amplitudes via shaking an optical lattice \cite{ShakenLatticePRLs}. They suggest that the anomalous counter-propagating modes may be identified using the momentum-resolved photoemission spectroscopy method of Ref.~\cite{PhotoemissionNature} which extracts a spectral function which in turn yields information on the number of states present for each energy and each momentum.

On the computational side, it is straightforward to extend our consideration to cases with more bands.  We have carried out calculations for cases with many Floquet bands and these suggest that the observations made in this work regarding the difference between DKL and KHL still hold. That is, if the number of bands is even, then there exist many edge modes with 0 or $\pi$ quasienergy values in the former model (DKL) but not in the latter (KHL); and if the number of bands is odd, there exist ACP chiral edge modes only in cases where the CS is of Type II mentioned under Eqs. (\ref{KHL-CS-OBC-ky}-\ref{DKL-CS-OBC-kx}). The DKL model for multiple band cases also shows both flat $0$ quasienergy modes as well as chiral modes, similar to those seen in the static context of Ref. \cite{DasSarmaZeroModesPRL2013}. Throughout this work, we have also viewed the system in a strictly stroboscopic manner. To be more precise, one may instead view the system continuously in time \cite{HighFreqBukov,NonStrobBukov}, which is beyond the scope of this work. Our stroboscopic treatment here should however be a fairly accurate representation of the physics, as our Floquet operators are both local in nature and do not transmit wavepackets over infinite distances within each period. Our results indicate that our observation here that the particular form of CS operator has a huge impact on the edge states is quite general. In future, it would be also interesting to study the implications of the particular form of other symmetry operators besides CS operators, such as time-reversal or particle-hole symmetry operators.

\appendix
\section{On the properties of CS operators of DKL and KHL}

We provide some mathematical details regarding the topological protection of the edge modes with $0$ and $\pi$ quasienergy values in the 2-band DKL model under open BCs along $x$. We denote the 2-band DKL Floquet operator under this BC simply as $U'_{\text{DKL}}(k_{y})$ for brevity. On the $0$ and $\pi$ quasienergy subspaces, $\Gamma_{\text{DK}}$ and $U'_{\text{DKL}}(k_{y})$ commute. This is easily seen as follows. Firstly, $U'_{\text{DKL}}(k_{y}) \Gamma_{\text{DK}}= \Gamma_{\text{DK}} U^{'\dagger}_{\text{DKL}}(k_{y}) $ due to the CS condition. Since any $0$ or $\pi$ quasienergy eigenstate of $U'_{\text{DKL}}(k_{y})$ is also an eigenstate of $U^{'\dagger}_{\text{DKL}}(k_{y})$ with the same eigenvalue, $\Gamma_{\text{DK}}$ and $U'_{\text{DKL}}(k_{y})$ thus commute within the $\omega=0$ and $\pi$ subspaces. The commutation enables us to choose the $0$ and $\pi$ quasienergy states to be common eigenstates of $\Gamma_{\text{DK}}$ and $U'_{\text{DKL}}(k_{y})$. Note that because $\left( \Gamma_{\text{DK}} \right)^{2} = \mathds{1}_{x}$, its only possible eigenvalues are $\pm 1$. This allows us to define two sublattices \cite{AsbothBBC} denoted $A$ and $B$, with projectors $\Pi_{A} = (1+\Gamma_{\text{DK}})/2$ and $\Pi_{B} = (1-\Gamma_{\text{DK}})/2$ respectively (i.e., sublattice A (B) consists of all the even (odd) sites). Each $0$ or $\pi$ quasienergy state then resides entirely on one lattice only. We denote such eigenstates as $\ket{\psi_{\omega,j}^{(\alpha)} }$, where $\omega = 0, \pi$, $\alpha \equiv A, B$ and $j$ is an index to label different states in the event that we have multiple eigenstates with the same quasienergy and same sublattice index $\alpha$.

Now, assume that at some $k_{y}$, we happen to have a number of $0$ and $\pi$ quasienergy states. Suppose we perturb the system in a way which preserves the chiral symmetry of Eq. (\ref{CS-DKL}). We may regard this increase as adding a CS-preserving perturbation to the original $\hat{H}'_{\text{eff}}(k_{y})$, defined by $U'_{\text{DKL}}(k_{y}) \equiv e^{-i\hat{H}'_{\text{eff}}(k_{y})}$. We denote this perturbation as $\hat{H}_{p}$. By definition of preservation of CS, it must be true that the anti-commutator $\{\Gamma_{\text{DK}},\hat{H}_{p}\}$ vanishes. Following Ref.~\cite{KitagawaNatComm}, we consider the following anti-commutator matrix element,
\begin{equation}
\bra{\psi_{\omega,j}^{(\alpha)}}\{\Gamma_{\text{DK}},\hat{H}_{p}\}\ket{\psi_{\omega ,j'}^{(\alpha ')}}=0.
\end{equation}
This tells us that, within the $0$ and $\pi$ quasienergy subspaces respectively, a CS-preserving perturbation $\hat{H}_{p}$ can only mix edge states living on different sublattices. This implies that the difference between the number of $0$ modes on the $A$ and $B$ sublattices must remain unchanged so long as the quasienergy gap remains open \cite{KitagawaNatComm}. Since varying $k_{y}$ by a small amount in $U'_{DKL}(k_{y})$ may be regarded as a CS-preserving perturbation, we expect to see $0$ and $\pi$ quasienergy states persist over a range of $k_{y}$ values so long as no gap-closing occurs. This is indeed seen in Figs. \ref{ordkr-vs-khm-2bands}(a) and \ref{ordkr-lattice-2band}.

We remind that the above analysis does not apply to the 2-band KHL model because the starting point of the analysis, which is the presence of CS in its Floquet operator $U'_{\text{KHL}}(k_{y})$ under open BCs along $x$, does not hold. This is the reason for the big difference in edge states between the 2-band DKL and KHL.

\section{Explicit Forms of $U''_{\text{DKL}} (k_{y})$ in the 2-band case}

The Floquet operator $U''_{\text{DKL}} (k_{y})$ in $\bar{k}_{x}$ representation and its effective Hamiltonian are written out below.
\begin{equation}
U''_{\text{DKL}}(k_{y})\equiv \sum _{\bar{k}_{x}} e^{-i \sum _{\bar{k}_{x}} H''_{\mathrm{eff}} (\bar{k}_{x},k_{y}) \otimes \ket{\bar{k}_{x}} \bra{\bar{k}_{x}}},
\end{equation}
where
\begin{eqnarray}
H''_{\mathrm{eff}}(\bar{k}_{x},k_{y}) &=& {\bf h}'' (\bar{k}_{x},k_{y}) \cdot {\bf \sigma}, \nonumber \\
{\bf h}'' (\bar{k}_{x},k_{y}) &=& E(\bar{k}_{x},k_{y}) {\bf n}'' (\bar{k}_{x},k_{y}), \nonumber \\
{\bf n}'' (\bar{k}_{x},k_{y}) &=& \left[ n_{x}'' (\bar{k}_{x},k_{y}), n_{y}'' (\bar{k}_{x},k_{y}), 0 \right], \nonumber \\
n_{x}'' (\bar{k}_{x},k_{y}) &=& \frac{\cos(\frac{\bar{k}_{x}}{2}) \cos(P) \sin(Q)  +\sin(\frac{\bar{k}_{x}}{2})\sin(P)}{\sin[E(\bar{k}_{x},k_{y})]}, \nonumber \\
n_{y}'' (\bar{k}_{x},k_{y}) &=& \frac{- \sin(\frac{\bar{k}_{x}}{2})\cos(P)\sin(Q) + \cos(\frac{\bar{k}_{x}}{2})\sin(P) }{\sin[E(\bar{k}_{x},k_{y})]},
\end{eqnarray}
and $E(\bar{k}_{x},k_{y}),P,Q$ are as they were defined previously.


\begin{thebibliography}{99}
\bibitem{credit} We note that Figs. \ref{khm-lattice-3band}(a),(c) and (e) are reproductions of Figs. 1(a), (b) and (d) of Ref. \cite{ZhaoKickedHall} using our own numerics in order to facilitate comparison with the DKL model.
\bibitem{lattice-rotor} The lattice sites $\ket{n_{x(y)}}$ here are mathematically equivalent to the discrete angular momentum eigenstates considered in the original ORDKR and KHM where a particle on a ring (i.e., a rotator space) was considered.
\bibitem{Quasienergies} To be precise, the quasienergy is defined as $\omega$ when the unitary eigenvalue of the Floquet operator is written as $e^{-i\frac{\omega T}{\hbar}}$, with $T$ being the time period. Since we always have the freedom to set $T$ and $\hbar$ to one, for brevity we will just write the unitary eigenvalue of the Floquet operator as $e^{-i\omega}$ and refer to $\omega$ as the quasienergy.
\bibitem{class-note} Although the 3-band KHL and DKL models possess CS, they belong in class A (and not AIII) because they possess an odd number of bands. It is clear from Sec.~2.2 of Ref.~\cite{Tenfoldway} that in their definition of CS, an added assumption of an even number of bands is made."
\end{thebibliography}
\end{document}